\RequirePackage{fix-cm}
\documentclass[smallextended,natbib]{svjour3}       
\smartqed  
\usepackage{graphicx}
\usepackage{subfig}
\usepackage{amsfonts}
\usepackage{amsmath}
\usepackage{multirow}
\usepackage{amssymb}
\usepackage[colorinlistoftodos, textwidth=4cm, shadow]{todonotes}

\begin{document}

\title{Automated Issue Assignment: Results and Insights from an Industrial Case
}

\titlerunning{Automated Issue Assignment: An Industrial Case}        

\author{Ethem Utku Aktas         \and
        Cemal Yilmaz 
}


\institute{Ethem Utku Aktas \at
              Softtech Inc., Research and Development Center, \\ 
              34947 Istanbul, Turkey \\
              \email{utku.aktas@softtech.com.tr}           
           \and
           Cemal Yilmaz \at
              Faculty of Engineering and Natural Sciences, \\
              Sabanci University, \\ 
              34956 Istanbul, Turkey \\
              \email{cyilmaz@sabanciuniv.edu}           
}

\date{Received: date / Accepted: date}

\maketitle

\begin{abstract}
{\em Softtech}, being a subsidiary of the largest private bank in Turkey, called {\em IsBank}, receives an average of $350$ issue reports from the field every day. Manually assigning the reported issues to the software development teams is costly and cumbersome. We automate the issue assignments using data mining approaches and share our experience gained by deploying the resulting system at Softtech/IsBank. Automated issue assignment has been studied in the literature. However, most of these works report the results obtained on open source projects and the remaining few, although they use commercial, closed source projects, carry out the assignments in a retrospective manner. We, on the other hand, deploy the proposed approach, which has been making all the assignments since Jan $12$, $2018$. This presents us with an unprecedented opportunity to observe the practical effects of automated issue assignment in the field and to carry out user studies, which  have not been done before in this context. We observe that it is not just about deploying a system for automated issue assignment, but also about designing/changing the assignment process around the system; the accuracy of the  assignments does not have to be higher than that of manual assignments in order for the system to be useful; deploying such a system requires the development of additional functionalities, such as detecting deteriorations in assignment accuracies in an online manner and creating human-readable explanations for the assignments; stakeholders do not necessarily resist change; and gradual transition can help stakeholders build confidence.
\end{abstract}
 
\section{Introduction}
\label{intro}

IsBank\footnote{https://www.isbank.com.tr} is the largest private bank in Turkey with $7.5$ million digital customers, $25$ thousand employees, $6566$ ATMs (Automated Teller Machines), and $1314$ domestic and $22$ foreign branches, providing a large variety of banking and financial services.

Softtech\footnote{https://softtech.com.tr} is an ISO-9001-certified subsidiary of IsBank and the largest software company of Turkey owned by domestic capital, providing  customer-oriented, business-critical solutions to IsBank by using universally-recognized lean techniques and agile processes with a diverse set of programming languages, platforms, and technologies, including COBOL, Java, C\#, C++, mainframe platforms, mobile/wearable platforms, security- and privacy-related technologies, natural language processing technologies, speech technologies, image/video processing technologies, and artificial intelligence technologies. 

When the wide range of software systems maintained by Softtech couple with the large user base owned by IsBank, who depend on these systems to carry out their day-to-day businesses, Softtech receives an average of $350$ issue reports from the field every day (around $90$ thousand reports per year). The reported issues range from bank clerks having software failures to bank customers facing software-related problems in any of the bank channels, including online, mobile, and ATM. Note that what we refer to as {\em issue reports} in this work are also often referred to as {\em bug reports} and {\em problem reports} in the literature. 

Most of the reported issues concern business-critical software systems. Therefore, both Softtech and IsBank need to handle these issues with utmost importance and urgency. To this end, two dedicated teams of employees are employed: one at IsBank with $50$ full-time clerks, called  {\em IT Help Desk} (IT-HD), and the other at Softtech with $30$ full-time employees, called {\em Application Support Team} (AST). IT-HD clerks collect the issues reported from the field, resolve the ones that they can (such as the issues addressed by basic troubleshooting guides), and if not, assign them to the software development teams at Softtech. AST members, on the other hand, being embedded in the development teams at Softtech, help the development teams manage the issue reports assigned to them.

An integral part of this process is to assign the issue reports to the right development teams, i.e., the ones that are responsible for resolving the reported issues. In the remainder of the paper, we refer to this task as {\em issue report assignment} (or {\em issue assignment}, in short). 


It turned out that the manual issue assignments carried out by the IT-HD clerks were costly and cumbersome, mainly due to the large number of issues received on a daily basis and the relatively large number of development teams that these issues should be assigned to (an average of $53$ development teams at any point given in time). Furthermore, incorrect assignments, were not only causing friction between IT-HD and AST, but also increasing the turnaround time for resolutions as the respective issue reports tented to bounce back and forth between IT-HD and AST until the right development team was located.

In this work, we automate the process of issue assignment by using data mining approaches and share our experience gained by deploying the resulting system, called {\em IssueTAG}, at IsBank and Softtech, which has been making all the issue assignments since its deployment on Jan $12$, $2018$.  

Automated issue assignment is indeed not a new idea \citep{murphy2004automatic, anvik2006should, wang2008approach, bhattacharya2012automated, jonsson2016automated, dedik2016automated}. Most of the existing works, however, report the results obtained on open source projects, such as Eclipse, Mozilla, and Firefox \citep{murphy2004automatic, anvik2006should, wang2008approach, bhattacharya2012automated}. Our work differs from these works in that we present an industrial case where we use the issue reports filed for commercial, closed-source software systems.

We, furthermore, assign issue reports to development teams, rather than to individual developers -- a decision we made based on our discussions with the IT-HD and AST teams. The former is more practical and realistic in industrial setups, because the latter does not take into account 1) the current workloads owned by the individual developers, 2) the changes in the team structures, such as the developers leaving or joining the teams, and 3) the current status of developers, such as the developers who are currently on leave of absence. Therefore, especially in the presence of close-knit development teams, which is the case with Softtech, assigning issue reports to the development teams help the teams make more educated decisions as to which team member the report should be addressed by. 


 
Moreover, rather than carrying out the issue assignments in the context of a single product, such as Eclipse, Mozilla, and Firefox, where the incoming issues are assigned to individual software engineers working on the product, we do the assignments at the level of an entire company (Softtech), which has $489$ software products comprised of around $100$ millions of lines of code (as of Feb 3, 2019). That is, we assign issue reports filed for any product owned by Softtech to the development teams responsible for resolving the reported issues. This is challenging because with the collection of software products maintained by Softtech, which heavily interact with each other in a business-critical environment by sharing many resources, such as databases, file systems, and GUI screens, the boundaries of the products from the perspective of issue reporting and management are not clear at all.

There are only few recent studies reporting the results obtained on closed-source, commercial software projects \citep{jonsson2016automated, dedik2016automated, lin2009empirical, helming2010automatic}. These studies, however, carry out the assignments in a retrospective and offline manner  by simply treating the actual issue databases as historical data. We have, on the other hand, deployed IssueTAG, which presented us with an unprecedented opportunity to observe the practical effects of automated issue assignment in practice as well as to carry out user studies, which (to the best of our knowledge) have not been done before in this context.

First, we observed that it is not just about deploying a data mining-based system for automated issue assignment, but also about designing/changing the assignment process around the system to get the most out of it. We, in particular, made simple, yet effective changes in the manual issue  assignment process employed at IsBank and Softtech  (Section~\ref{inaction}).

Second, the accuracy of the assignments does not have to be higher than that of manual assignments in order for the system to be useful, which is further validated by the user studies we carried out on actual stakeholders in the field (Section~\ref{userEvals}). In a nutshell, although the daily assignment accuracy of IssueTAG was slightly lower than that of manual assignments ($0.831$ vs. $0.864$), it reduced the manual effort required for the assignments by about $5$ person-months per year and improved the turnaround time for resolving the reported issues by about $20$\% (Section~\ref{deployment:analysis}). Furthermore, about $79$\% of the stakeholders participated in our user study ``agreed'' or ``strongly agreed'' that the system was useful (Section~\ref{userEvals}).

Third, we observed that deploying a data mining-based approach for automated issue assignments, requires the development of additional functionalities, which are not necessarily foreseen before the deployment. We have, in particular, developed two additional functionalities, both of which, to the best of our knowledge, have not been evaluated before in the context of issue assignment. One functionality we needed was to monitor the assignment accuracy of the system and detect deteriorations in an online manner, so that corrective actions, such as recalibrating the models, can be taken in time. To this end, we have developed a {\em change point detection}-based approach (Section~\ref{monitor}). Another functionality we needed, which we also did not foresee before the deployment of IssueTAG, was to create human-readable, non-technical explanations for the assignments made by the system. This was indeed a need we came to realize when we received several phone calls from the stakeholders shortly after the deployment of IssueTAG, demanding explanations as to why certain issue reports (especially, the incorrect-assigned ones) were assigned to them.  Note that this is not a trivial task at all, especially when the underlying data mining models are not human readable. To this end, we have generated model-agnostic explanations \citep{ribeiro2016should} and carried out a user study to evaluate the quality of these explanations (Section~\ref{monitor}).

Last but not least, we observed that stakeholders do not necessarily resist change. In particular, we did not receive any objection at all to the deployment of IssueTAG. We believe that this was because all the stakeholders believed that they would benefit from the new system and none of them felt threatened by it (Section~\ref{lessons}). We, furthermore, observed that gradual transition helped stakeholders build confidence in IssueTAG, which, in turn, facilitated the acceptance of the system  (Section~\ref{lessons}).

\pagebreak 

More specifically, the research question we address in this work are:

\begin{list}{-}{}

\item RQ1: How do the existing approaches for automated issue assignment compare with each other?

\item RQ2: How do the amount and time locality of training data affect the assignment accuracies?

\item RQ3: How does automated issue assignment compare to manual issue  assignment in practice?

\item RQ4: Is IssueTAG perceived as useful by the end-users?

\item RQ5: Can the issue assignments made by the underlying data mining model be explained in a non-technical manner?

\item RQ6: Can the deteriorations in the assignment accuracies be automatically detected in an online manner?
 
\end{list}

The remainder of the paper is organized as follows: Section~\ref{caseDescription} describes the issue assignment process employed at IsBank and Softtech before the deployment of IssueTAG; Section~\ref{existing} evaluates the existing approaches for automated issue assignment on the collection of issue reports  maintained by IsBank and Softtech (RQ1); Section~\ref{timeLocality} presents the empirical studies we carried out to determine the amount and time locality of the training data required for training/recalibrating the underlying data mining models(RQ2); Section~\ref{inaction} deploys IssueTAG and evaluates its effects in practice (RQ3); Section~\ref{userEvals} carries out a user study on the end users of IssueTAG to evaluate whether the deployed system is perceived as useful (RQ4); Section~\ref{explain} presents an approach for automatically generating explanations for the assignments and evaluates it by conducting a user study (RQ5); Section~\ref{monitor} describes and evaluates a change point detection-based approach for detecting deteriorations in assignment accuracies (RQ6); Section~\ref{lessons} presents lessons learnt; Section~\ref{threats} discusses threats to validity; Section~\ref{related} presents related work; and Section~\ref{conclusion} concludes with potential avenues for future work.
 
\section{Case Description}
\label{caseDescription}

At IsBank and Softtech, the issue reports, as they typically concern business-critical systems, are handled with utmost importance and urgency. To this end, two dedicated teams are employed, the sole purpose of which is to manage the reported issues, namely IT-HD (IT Help Desk) and AST (Application Support Team).


\subsection{IT Help Desk} 

The IT-HD team is employed at IsBank and it consists of $50$ full-time, (mostly) non-technical clerks, who are internally referred to as Level 1 employees, indicating the level of technical competency they have. When a bank employee or a bank customer faces with an IT-related issue, they call IT-HD on the phone. The IT-HD clerk listens to the issue, collects the details as needed, records them, and resolves the reported issue right away if it is an issue that can be resolved by an IT-HD clerk, such as the ones documented in basic troubleshooting guides. If not, the clerk is responsible for dispatching the issue to the proper entity/unit in the company. In the case of a software-related issue, the clerk files an issue report to Softtech. 

\subsection{Issue Reports} 

An issue report, among other information, such as the date and time of creation, has two parts: a one-line summary and a description, both of which are written in Turkish. The former captures the essence of the issue, whereas the latter describes the issue, including the expected and observed behavior of the system, and provides information to reproduce the reported issue \citep{bettenburg2008makes}. Note that the aforementioned issue reports do not have any field conveying categorical information, such as product, component, and version information. The reason is that since the collection of software products maintained by Softtech are heavily interacting with each other in a  business-critical environment, sharing many resources, such as databases, file systems, and GUI screens, the boundaries of the products/components from the perspective of issue reporting and management are not clear at all. For example, a single GUI screen can have multiple tabs, each of which is maintained by a different development team. A single tab can, in turn, have a number of widgets, each of which is under the responsibility of a different team. Almost all of the GUI screens interact with the core banking system, which is maintained by a different set of development teams. The core can be accessed via different banking channels, such as online, mobile, ATM, and SMS (Short Message Service), each of which has a dedicated set of development teams. Last but not least, financial transactions are typically carried out by using multiple GUI screens, widgets, and channels, crossing the boundaries of multiple development teams.
 
\subsection{Application Support Team (AST)}

The AST team is employed at Softtech and it consists of $30$ full-time, Level 2 employees. That is, in terms of technical competency, the AST employees are somewhere between Level 1 IT-HD clerks and Level 3 software engineers. AST employees are embedded in development teams, which are consisted of software engineers. The same AST member can work with multiple development teams and a development team can have multiple AST members. The sole responsibility of an AST member embedded in a development team is to manage the collection of issue reports assigned to the team. When a new issue report is assigned to a development team, the AST member embedded in the team is typically the first one to examine the report. If the AST member can resolve the reported issue, he/she first resolves it and then closes the report on behalf of the team. Otherwise, the AST member notifies the development team about the newly reported issue by, for example, assigning it to a software engineer in the team or by creating a task for the team and linking it to the issue report. Note that AST members, although they are not considered to be software engineers, can still resolve some of the reported issues as not all of these issues may require changes in the code base. Some issues, for example, are resolved by running pre-existing scripts, which can automatically diagnose and fix the problems or by manually updating certain records in the database. Therefore, the ultimate goal of the AST members is to reduce the workload of software engineers by resolving the issues that do not require code changes.

\begin{figure}
  \includegraphics[width=1\textwidth]{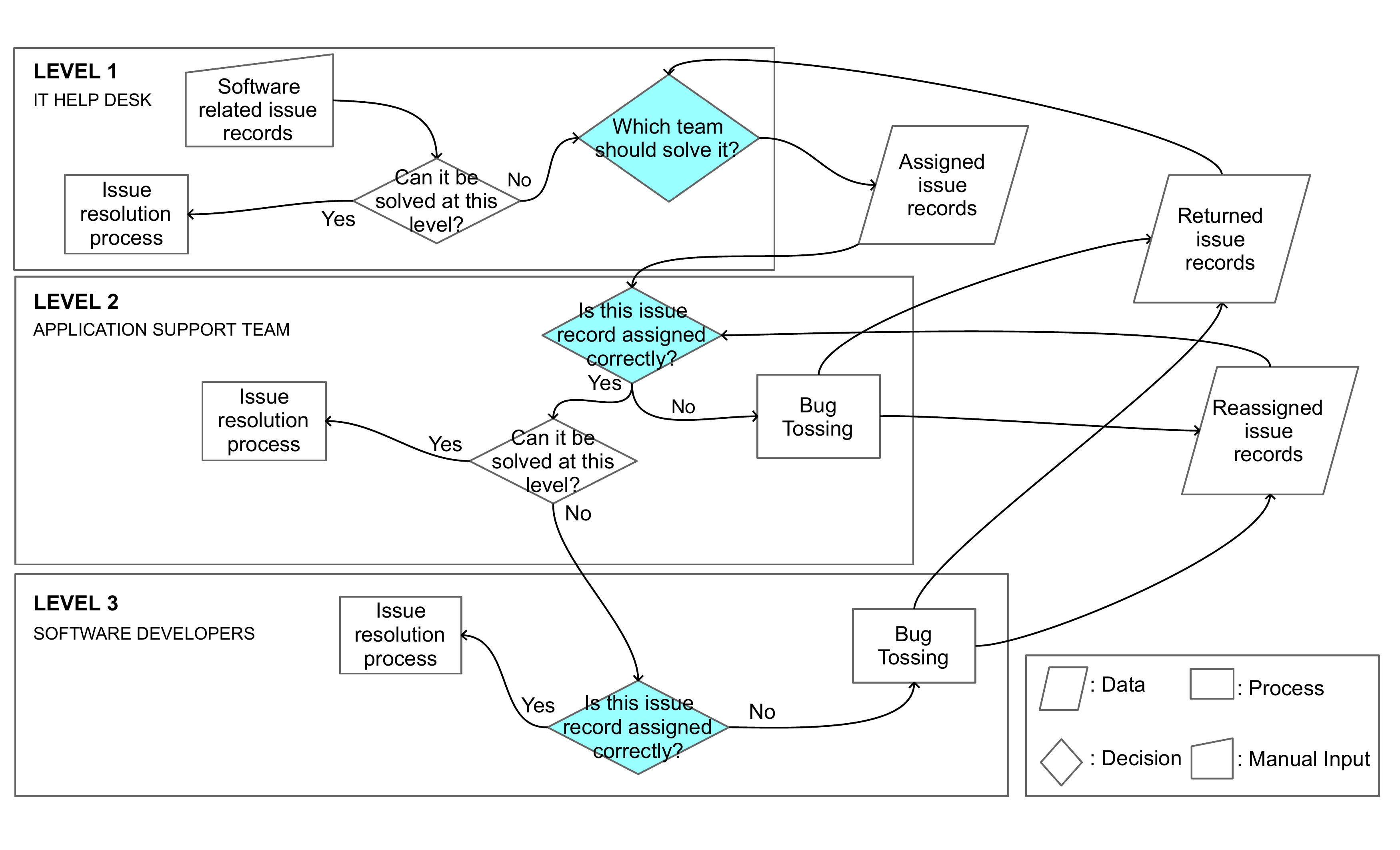}
\caption{Issue report assignment process before the deployment of IssueTAG.}
\label{xprocess}   
\end{figure}

\subsection{Manual Issue Assignment Process} 

Before the deployment of IssueTAG, IT-HD clerks, after creating an issue report, was assigning it to a development team. To this end, they were maintaining a knowledge base, which was simply comprised of spreadsheets mapping certain keywords with development teams. In the presence of an incorrect assignment, although the AST member(s) or the software engineers in the respective development team could reassign the issue  to a different team, the incorrectly assigned reports were often returned back to IT-HD for reassignment. Figure~\ref{xprocess} summarizes the assignment process. The issue reports are managed by using Maximo\footnote{https://www.ibm.com/products/maximo} at IsBank and by using Jira\footnote{https://www.atlassian.com/software/jira} at Softtech.

\subsection{Issues with the Manual Assignment Process}

There were a number of issues with the aforementioned process. First, the learning curve for the IT-HD clerks (especially for the new hires) for excelling in team assignments was generally steep due to the large number of issue reports received on a daily basis (an average of about $350$ issue reports) and the relatively large number of products and development teams present (more than $450$ products and between $47$ and $57$ teams at any given point in time). Second, although IT-HD clerks were using a knowledge base to help with the assignments, it was maintained in an ad hoc manner, which was error prone, cumbersome, and time consuming. Last but not least, incorrect assignments were not only causing frictions between the IT-HD clerks and the AST members, but also increasing the turn around time for resolutions as the incorrectly-assigned issue reports would typically bounce back and forth between the IT-HD clerks and AST members ({\em bug tossing}) until the correct development team was located, which was causing a great deal of wasted time.

\section{Evaluating Existing Issue Assignment Approaches}
\label{existing}

IsBank and Softtech wanted to improve their current practices. To this end, we, in this work, automate the issue assignments.

We start with investigating our first research question RQ1: ``How do existing approaches for automated issue assignment compare with each other?'' This is important as the results of this study will determine the approach to be used in the deployed system.


Note that our goal in this work is neither to propose yet another approach for automated issue assignment nor to evaluate all existing approaches to determine the best possible approach, but to identify an existing approach that can produce similar or better assignment accuracies with the manual assignment process employed at IsBank/Softtech and that can be developed and deployed with as little risk as possible. After all, most of the issue reports the system will process, concern business-critical software systems. Therefore, neither IsBank nor Softtech was willing to take too much risk.


\subsection{Approach}
\label{existing:approach}

To carry out the study, we have determined a number of approaches, which had been shown to be effective for automated issue assignment \citep{murphy2004automatic, anvik2006should, bhattacharya2012automated, anvik2011reducing, jonsson2016automated}. We then empirically evaluated them by using the issue database, which has been maintained by Softtech since December 2016. 

In particular, we cast the problem of issue assignment to a classification problem where the natural language descriptions in issue reports are analyzed by using various classification algorithms.


\subsubsection{Representing Issue Reports}
\label{existing:representation}

Given an issue report, we first combine the ``description'' and ``summary'' parts of the report, then tokenize the combined text into terms, and finally remove the non-letter characters, such as punctuation marks, as well as the stop words, which are extremely common words of little value in classifying issue reports \citep{manning2010introduction}, such as ``the'', ``a'', and ``is.'' We opt not to apply stemming in this work as an earlier work suggests that stemming has a little effect (if any at all) in issue assignments \citep{murphy2004automatic}, which is also consistent with the results of our initial studies where stemming slightly reduced the assignment accuracies. 

We then represent an issue report as an $n$-dimensional vector. Each element in this vector corresponds to a term and the value of the element depicts the weight (i.e., ``importance'') of the term for the  report. The weights are computed by using the well-known {\em tf-idf} method \citep{manning2010introduction}.


The $\operatorname{tf-idf}$ method combines two scores: term frequency ({\em tf}) and inverse document frequency ({\em idf}). For a given term $t$ and an issue report $r$, the term frequency {\em tf}$_{t, r}$ is the number of times $t$ appears in $r$. The more $t$ appears in $r$, the larger {\em tf}$_{t, r}$ is. The inverse document frequency of $t$ ($idf_t$), on the other hand, is: 

\begin{equation}
idf_t=log(\frac{N}{df_t}),
\end{equation}

\noindent where $N$ is the total number of issue reports and $df_t$ is the number of issue reports, in which $t$ appears. The fewer the issue reports $t$ appears in, the larger $idf_t$ is.

Given {\em tf}$_{t, r}$ and $idf_t$, the $\operatorname{tf-idf}$ score of the term $t$ for the issue report $r$ is computed as follows:

\begin{equation}
\operatorname{tf-idf}_{t, r}= tf_{t, r} * idf_t.
\end{equation}

\noindent Consequently, the more a term $t$ appears in an issue report $r$ and the less it appears in other issue reports, the more important $t$ becomes for $r$, i.e., the larger $\operatorname{tf-idf}_{t, r}$ is.
 
\subsubsection{Issue Assignments}
\label{existing:assignment}

Once an issue report is represented as an ordered vector of $\operatorname{tf-idf}$ scores, the problem of assignment is cast to a classification problem where the development team, to which the issue report should be assigned, becomes the class to be predicted and the $\operatorname{tf-idf}$ scores of the report become the attributes, on which the classification will be based on.

We train two types of classifiers: {\em level-0 classifiers}, each of which is comprised of an individual classifier, and {\em level-1 classifiers}, which were obtained by combining multiple level-0 classifiers using stacked generalization -- an ensemble technique to combine multiple individual classifiers \citep{wolpert1992stacked}. All the classifiers we experiment with in this study have been shown to be effective for automated issue assignment \citep{murphy2004automatic, anvik2006should, bhattacharya2012automated, anvik2011reducing, jonsson2016automated}.

For the level-0 classifiers, we use multinomial naive bayesian \citep{manning2010introduction}, decision tree \citep{breiman2017classification}, k-nearest neighbor \citep{manning2010introduction}, logistic regression \citep{bishop2006pattern}, random forest \citep{breiman2001random}, and linear support vector classifiers (SVCs) \citep{joachims1998text}. 

For the level-1 classifiers, we first train and evaluate our level-0 classifiers by using the same training and test sets for each classifier. We then use the prediction results obtained from these level-0 classifiers to train a level-1 classifier, which combines the probabilistic predictions of the level-0 classifiers using linear logistic regression \citep{wolpert1992stacked}.

Inspired from~\citep{jonsson2016automated}, we, in particular, train two types of level-1 classifiers: {\em BEST} and {\em SELECTED}. The BEST ensemble is comprised of $k$ (in our case, ${k=\{3, 5\}}$) level-0 classifiers with the highest assignment accuracies, where as the SELECTED ensemble is comprised of a diversified set of $k$ (in our case, ${k=\{3, 5\}}$) level-0 classifiers, i.e., the ones with different representation and classification approaches, which are selected regardless of their classification accuracies, so that errors of individual classifiers can be averaged out by better spanning the learning space \citep{wolpert1992stacked}. Note that the BEST and SELECTED ensembles are not necessarily the same because the best performing level-0 classifiers may not be the most diversified set of classifiers. More information on how these ensembles are created can be found in Section~\ref{existing:evaluation}.


Furthermore, for the baseline classifier, which we use to estimate the baseline classification accuracy for our classifiers, we assign all issue reports to the team that have been assigned with the highest number of issue reports. That is, our baseline classifier always returns the class with the highest number of instances as the prediction.

\subsection{Evaluation}
\label{existing:evaluation}

We have conducted a series of experiments to evaluate the assignment accuracies of the level-0 and level-1 classifiers.

\subsubsection{Experimental Setup}
\label{existing:setup}

In these experiments, we used the issue reports submitted to Softtech between June $1$, $2017$ and November $30$, $2017$ as the training set and the issue reports submitted in the month of December 2017 as the test set. We picked this time frame because it provided us with a representative data set in terms of the number of issue reports submitted, the number of teams present, and the distribution of the reported issues to these teams.

For the aforementioned time frame, we had a total number of $51,041$ issue reports submitted to $65$ different teams. Among all the issue reports of interest in this section as well as in the remainder of the paper, we only used the ones that were marked as ``closed,'' indicating that the reported issues had been validated and resolved. Furthermore, as the correct assignment for an issue report, we used the development team that had closed the report. The remainder of the issue reports were ignored as it was not yet certain whether these reports were valid or whether the development teams, to which they were currently assigned, were correct. After this filtering, a total of $47,123$ issue reports submitted to $64$ different development teams remained for analysis in this study.

To create the level-1 classifiers, we combined $3$ or $5$ individual classifiers, i.e., ${k=3}$ or ${k=5}$. We used the latter setting as it was also the setting used in a recent work \citep{jonsson2016automated}. We used the former setting as it was the best setting we could empirically determine for ensemble learning, i.e., the one that produced the best assignment accuracies. In the remainder of the paper, these models are referred to as {\em BEST-3}, {\em SELECTED-3}, {\em BEST-5}, and {\em SELECTED-5}.


The BEST-3 and BEST-5 models were obtained by combining Linear SVC-Calibrated, Logistic Regression, and K-Neighbours; and Linear SVC-Calibrated, Logistic Regression, K-Neighbours, Random Forest, and Decision Tree classifiers, respectively, as these were the classifiers providing the best assignment accuracies. The SELECTED-3 and SELECTED-5 models, on the other hand, were obtained by combining Linear SVC-Calibrated, K-Neighbours, and Multinomial Naive Bayesian; and Linear SVC-Calibrated, Logistic Regression, K-Neighbours, Random Forest, and Multinomial Naive Bayesian classifiers, respectively, with the goal of better spanning the learning space by increasing the diversity of the classification algorithms ensembled. Note further that to include SVCs in level-1 classifiers, we used calibrated linear SVCs instead of linear SVCs as we needed to have class probabilities to ensemble individual classifiers \citep{ting1999issues}, which are not supported by the latter.

The classifiers were trained and evaluated by using the {\tt scikit-learn} Python library (for level-0 classifiers) \citep{pedregosa2011scikit} and {\tt mlxtend} (for level-1 classifiers) \citep{raschka2018mlxtend} packages. All of the classifiers (unless otherwise stated) were configured with the default settings and the experiments were carried out on a dual-core Intel(R) Xeon(R) E5-2695 v4 2.10 GHz machine with 32 GB of RAM running Windows Server 2012 R2 as the operating system.

\subsubsection{Evaluation Framework}
\label{existing:evalframework}

To evaluate the quality of the assignments obtained from different classifiers, we used well-known metrics, namely {\em accuracy} and weighted {\em precision}, {\em recall}, and {\em F-measure} \citep{manning2010introduction}. Accuracy, which is also referred to as {\em assignment accuracy} in the remainder of the paper, is computed as the ratio of correct issue assignments. Precision for a particular development team (i.e., class) is the ratio of the issue reports that are correctly assigned to the team to the total number of issue reports assigned to the team. Recall for a team is the ratio of the issue reports that are correctly assigned to the team to the total number of issue reports that should have been assigned to the team. F-measure is then computed as the harmonic mean of precision and recall, giving equal importance to both metrics. Note that each of these metrics takes on a value between $0$ and $1$ inclusive. The larger the value, the better the assignments are. Furthermore, we report the results obtained by both carrying out $10$-fold cross validation on the training data and carrying out the analysis on the test set. 

To evaluate the cost of creating the classification models, we measured the time it took to train the models. The smaller the training time, the better the approach is. 


\begin{table}
\small
\caption{Accuracy (A) and weighted precision (P), recall (R), and F-measure (F) values obtained from different classification models as well as the training times of these models.}
\label{tbl:existing}      
\resizebox{\columnwidth}{!}   
{\begin{tabular}{|l|ll|llll|}
\hline
& \multicolumn{2}{|c|}{using training set with} & \multicolumn{4}{|c|}{using}  \\
& \multicolumn{2}{|c|}{10-fold cross validation} & \multicolumn{4}{|c|}{test set} \\
\hline
\multirow{2}{*}{classifier} & & training &  & & &  \\
& A & time & P & R & F & A \\
\hline
Baseline & 0.10 & - & 0.01 & 0.12 & 0.03 & 0.12 \\
Multinomial NB & 0.47 (+/- 0.01) & 31 s & 0.70 & 0.52 & 0.50 & 0.52 \\
Decision Tree & 0.66 (+/- 0.02) & 50 m 11 s & 0.64 & 0.63 & 0.63 & 0.63 \\
K-Neighbours & 0.73 (+/- 0.02) & 1 m 4 s & 0.71 & 0.72 & 0.71 & 0.72 \\
Logistic Regression & 0.74 (+/- 0.01) & 18 m 37 s & 0.76 & 0.74 & 0.74 & 0.74 \\
Random Forest & 0.66 (+/- 0.02) & 51 m 43 s & 0.64 & 0.65 & 0.63 & 0.65 \\
Linear SVC & \textbf{0.82 (+/- 0.01)} & 3 m 32 s & \textbf{0.80} & \textbf{0.80} & \textbf{0.80} & \textbf{0.80} \\
Linear SVC-Calibrated & 0.81 (+/- 0.01) & 7 m 50 s & \textbf{0.80} & 0.79 & 0.79 & 0.79 \\
\hline
BEST-5 & 0.67 (+/- 0.02) & 2 h 20 m 38 s & 0.65 & 0.64 & 0.64 & 0.64 \\
SELECTED-5 & 0.80 (+/- 0.01) & 1 h 49 m 11 s & 0.79 & 0.78 & 0.78 & 0.78 \\
\hline
BEST-3 & 0.81 (+/- 0.01) & 56 m 7 s & \textbf{0.80} & 0.79 & 0.79 & 0.79 \\
SELECTED-3 & 0.81 (+/- 0.01) & 32 m 57 s & \textbf{0.80} & 0.79 & 0.79 & 0.79 \\
\hline
\end{tabular}}
\end{table}

\subsubsection{Data and Analysis}
\label{existing:analysis}

Table~\ref{tbl:existing} summarizes the results we obtained. We first observed that all the classifiers we trained performed better than the baseline classifier. While the baseline classifier provided an accuracy of $0.10$ on the training set and $0.12$ on the test set, those of the worst-performing classifier were $0.47$ and $0.52$, respectively.


We then observed that the SELECTED ensembles generally performed similar or better than the BEST ensembles, supporting the conjecture that using diversified set of classifiers in an ensemble can help improve the accuracies by better spanning the learning space. For example, while the accuracy of the BEST-5 ensemble was $0.67$ on the training set and $0.64$ on the test set, those of the SELECTED-5 ensemble were $0.80$ and $0.78$, respectively. Furthermore, the ensembles created by using $3$ level-0 classifiers, rather than $5$ level-0 classifiers, performed slightly better on our data set. For example, while the accuracy of the SELECTED-5 ensemble was $0.80$ on the training set and $0.78$ on the test set, those of the SELECTED-3 ensemble were $0.81$ and $0.79$, respectively.

Last but not least, among all the classifiers, the one that provided the best assignment accuracy (as well as the best F-measure) and did so at a fraction of the cost, was the linear SVC classifier (Table~\ref{tbl:existing}). While the linear SVC classifier provided an accuracy of $0.82$ on the training data set and $0.80$ on the test set with a training time of about three minutes, the runner-up classifiers, namely the SELECTED-3 and BEST-3 ensembles, provided the accuracies of $0.81$ and $0.79$, respectively, with a training time of about half an hour or more.

Based on both the assignment accuracies and the costs of training obtained from various classifiers using our data set, we have decided to employ linear SVC in IssueTAG. Consequently, all of the results presented in the remainder of the paper were obtained by using linear SVC classifiers. 
 

\section{Time Locality and Amount of Training Data}
\label{timeLocality}

In this section, we investigate our second research question (RQ2): ``How do the amount and time locality of training data affect the assignment accuracies?'' Note that the results of this study will be used to determine the amount of training data (e.g., how many issue reports we should use) as well as the time locality of this data (i.e., how much back in time we should go) required for preparing the training data every time the underlying classification model needs to be retrained.


\subsection{Approach}
\label{timeLocality:approach}

To answer these questions, we use the {\em sliding window} and {\em cumulative window} approaches introduced in~\citep{jonsson2016automated}. More specifically, we conjecture that using issue reports from ``recent past'' to train the prediction models, as opposed to using the ones from ``distant past'', can provide better assignment accuracies since organizations, products, teams, and issues may change overtime.

To evaluate this hypothesis, we take a long period of time $T$ (in our case, $13$ months) and divide it into a consecutive list of calendar months ${T=[m_1, m_2, \dots]}$. For every month $m_i \in T$, we train and evaluate a linear SVC model. To this end, we use all the issue reports submitted in the month of $m_i$ as the test set and all the issue reports submitted in the month of $m_j$ as the training set, where $i - j = \Delta$, i.e., the sliding window approach in~\citep{jonsson2016automated}. Note that given $m_i$ and $\Delta$, $m_j$ is the month, which is $\Delta$ months away from $m_i$ going back in time. For every month $m_i \in T$, we repeat this process for each possible value of $\Delta$ (in our case, $\Delta \in \{1, \dots, 12\}$). By fixing the test set and varying the training sets, such that they come from different historical periods, we aim to measure the effect of time locality of the training data on the assignment accuracies.

\begin{figure}
  \includegraphics[width=1\textwidth]{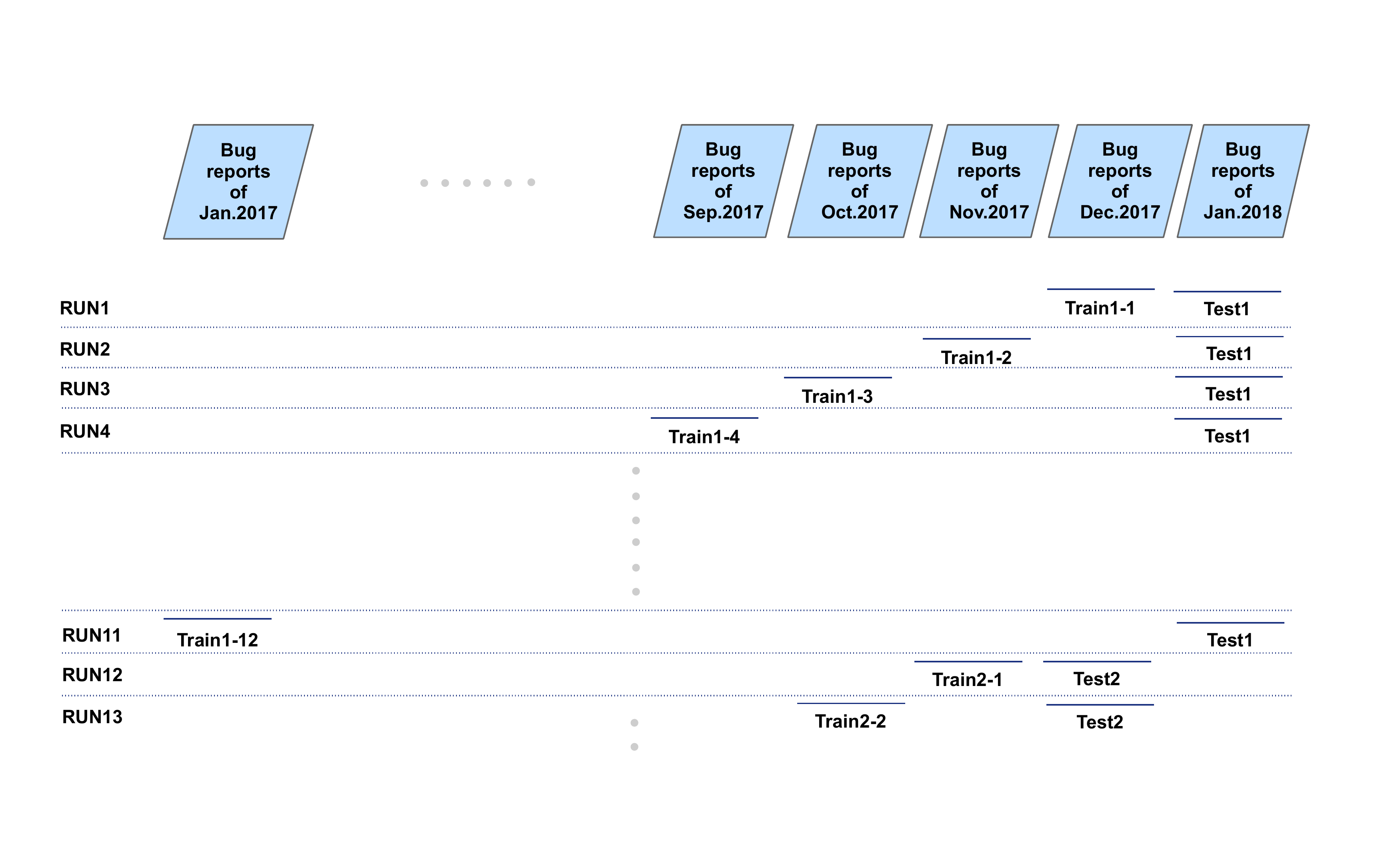}
\caption{Overview of the sliding window approach to study the effect of the time locality of training data on assignment accuracies.}
\label{fig:slidingWindow}
\end{figure}

Figure~\ref{fig:slidingWindow} illustrates the sliding window approach using the period of time from Jan $1$, $2017$ to Jan $31$, $2018$. For example, for the month of Jan $2018$, we train a total of $12$ classification models, each of which was trained by using all the issue reports submitted in a distinct month of $2017$ (marked as Train1-1, Train1-2, $\dots$, Train1-12) and separately test these models using all the issue reports submitted in the month of Jan, $2018$ as the test set (marked as Test1). We then repeat this process for every month in the time period of interest, except for Jan $2017$ as it does not have any preceding months. That is, for Dec $2017$ (marked as Test2), we train and evaluate $11$ models (marked as Train2-1, Train2-2, $\dots$), for Nov $2017$, we train and evaluate $10$ models, etc.


To evaluate the effect of the amount of training data on the assignment accuracies, we use a related approach, called the  cumulative window approach \citep{jonsson2016automated}. This approach, as is the case with the sliding window approach, divides a period of interest $T$ in to a consecutive list of months ${T=[m_1, m_2, \dots]}$. Then, for every possible pair of $m_i \in T$ and $\Delta$, we train and evaluate a classification model, where all the issue reports submitted in the month of $m_i$ are used as the test set and all the issue reports submitted in the preceding $\Delta$ months, i.e., $\{m_j \in T \mid 1 \leq i-j \leq \Delta\}$, are used as the training set.

\begin{figure}
  \includegraphics[width=1\textwidth]{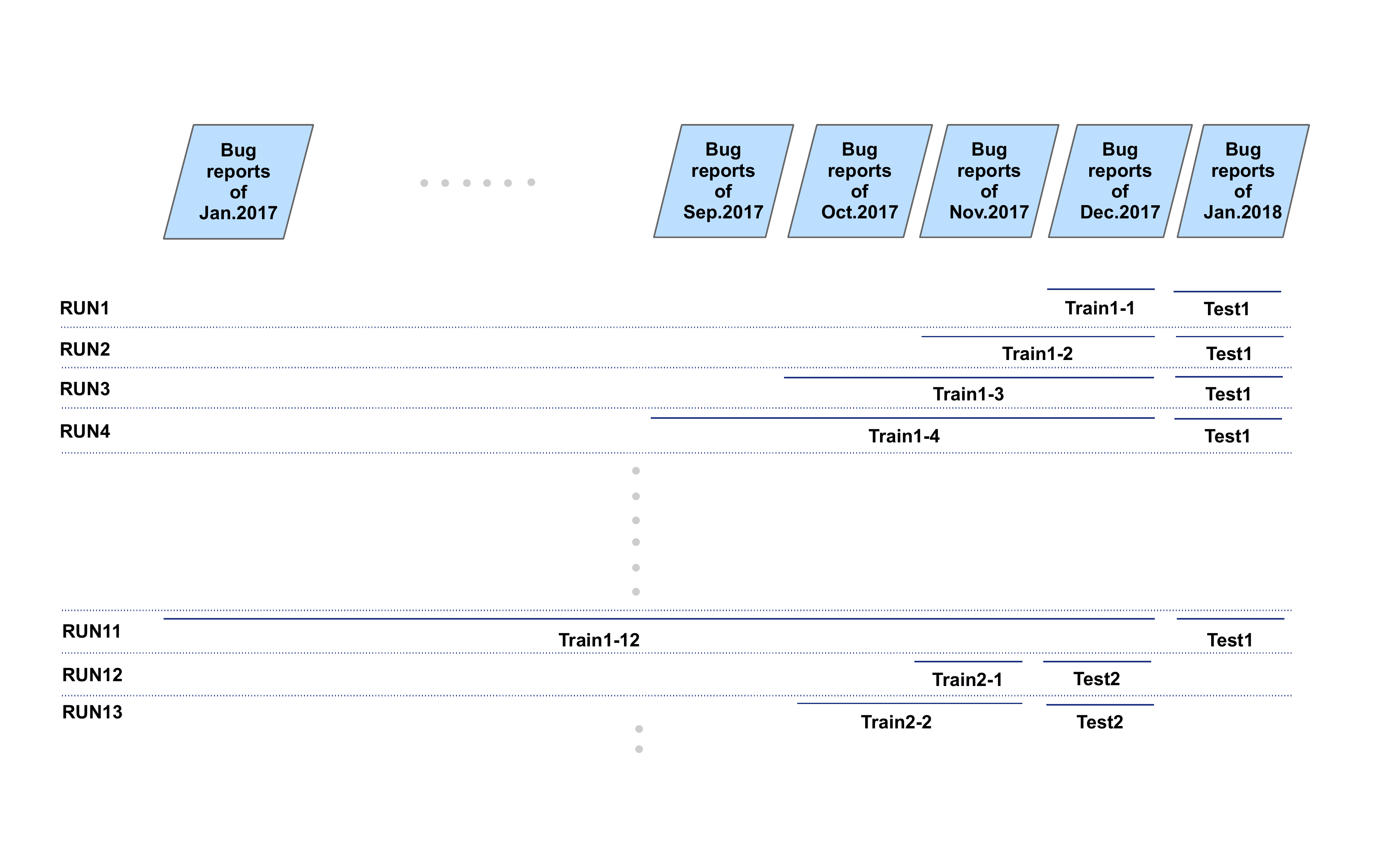}
\caption{Overview of the cumulative window approach to study the effect of the amount of training data on assignment accuracies.}
\label{fig:cumulativeWindow}     
\end{figure}

Figure~\ref{fig:cumulativeWindow} illustrates the approach. For example, for the month of Jan $2018$, we train a total of $12$ classification models. The first model is created by using the previous month's data (marked as Train1-1), the second model is created by using the previous two months' data  (marked as Train1-2), and the last model is created by using the previous year's data  (marked as Train1-12). The same process is repeated for every possible month in the period of interest.

\subsection{Evaluation}
\label{timeLocality:evaluations}
 
We conducted a series of experiments to evaluate the effect of the amount and time locality of training data on assignment accuracies.

\subsubsection{Experimental Setup}
\label{timeLocality:setup}

In these experiments, we used all the issue reports that were submitted during the period from Jan $1$, $2017$ to Jan $31$, $2018$. The summary statistics for this data set can be found in Table~\ref{tbl:data}. All told, we have trained and evaluated a total of $144$ linear SVC models for this study. All the experiments were carried out on the same platform with the previous study (Section~\ref{existing:setup}).

\begin{table}[t]
\centering
\caption{Number of issue reports submitted.}
\label{tbl:data}     
\begin{tabular}{|r|c|c|}
\hline
      & \# of issue reports & \# of teams \\
month & submitted            & assigned \\

\hline
Jan 2017 & 6364 & 57 \\
Feb 2017 & 5038 & 56 \\
Mar 2017 & 7188 & 57 \\
Apr 2017 & 6623 & 55 \\
May 2017 & 6601 & 56 \\
Jun 2017 & 6145 & 56 \\
Jul 2017 & 6341 & 53 \\
Aug 2017 & 6025 & 54 \\
Sep 2017 & 5961 & 54 \\
Oct 2017 & 6774 & 52 \\
Nov 2017 & 7996 & 54 \\
Dec 2017 & 7881 & 49 \\
Jan 2018 & 7426 & 51 \\
\hline
Total & 86363 & 69 \\
\hline
\end{tabular}
\end{table}

\subsubsection{Evaluation Framework}
\label{timeLocality:evalframework}

We used the assignment accuracies (Section~\ref{existing:evalframework}) for evaluations.


\subsubsection{Data and Analysis}
\label{timeLocality:analysis}

\begin{figure*}
  \includegraphics[width=1\textwidth]{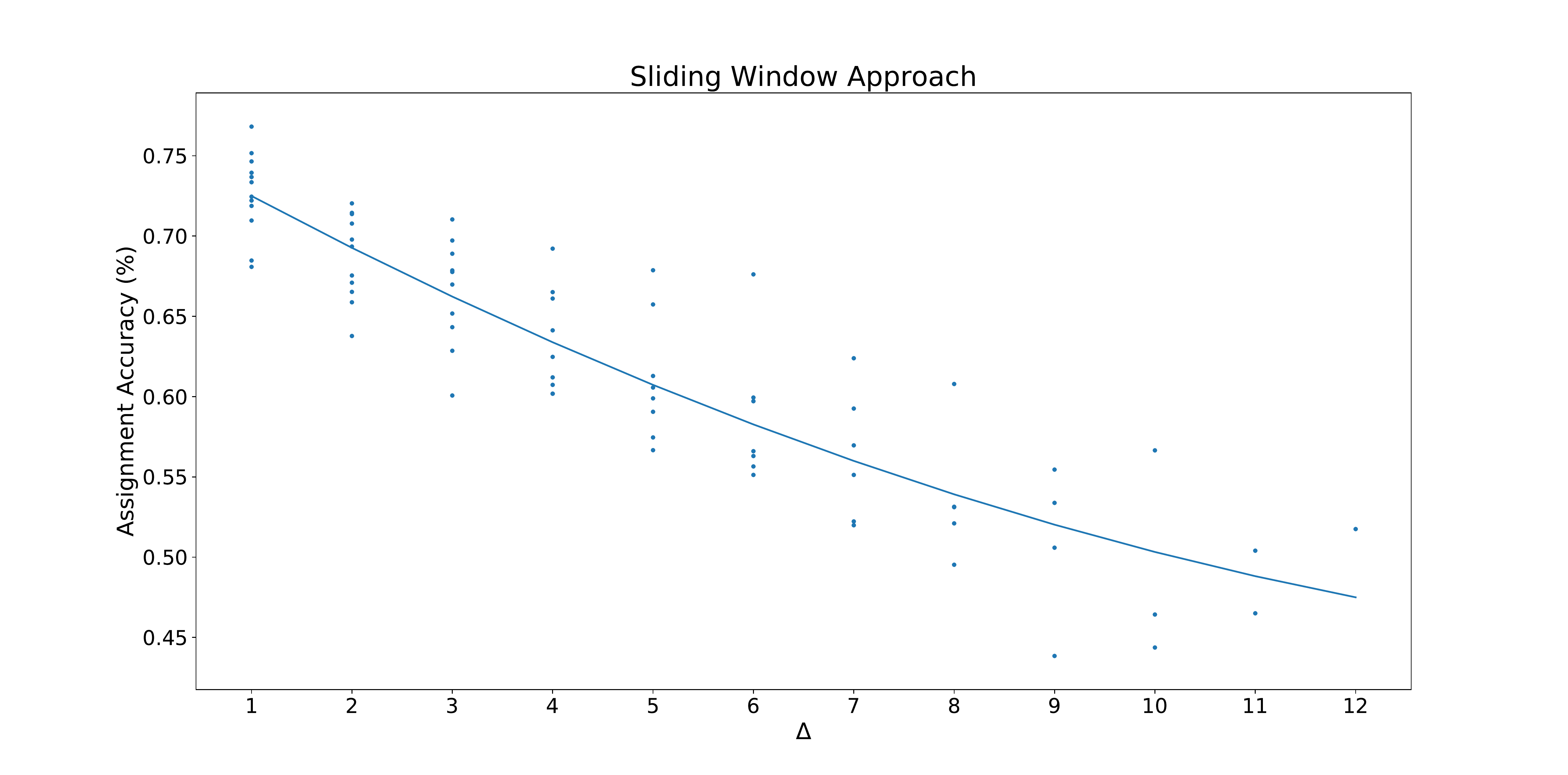}
\caption{Assignment accuracies obtained from the sliding window approach.}
\label{plot:slidingWindow}     
\end{figure*}

\begin{figure*}
  \includegraphics[width=1\textwidth]{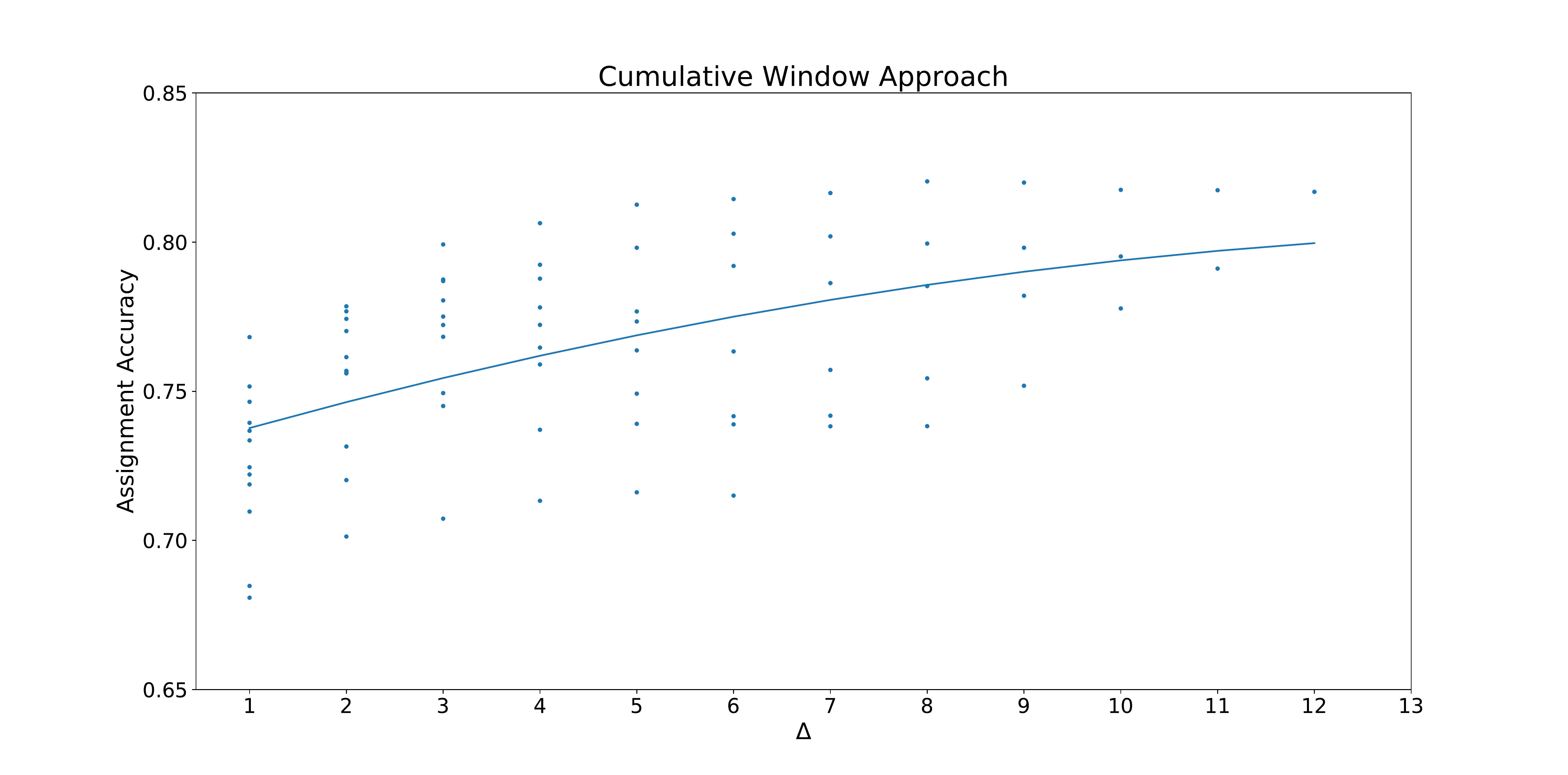}
\caption{Assignment accuracies obtained from the cumulative window approach.}
\label{plot:cumulativeWindow}      
\end{figure*}

Figures~\ref{plot:slidingWindow} and~\ref{plot:cumulativeWindow} represent the results we obtained from the sliding window and cumulative window approach, respectively. In these figures, the vertical and horizontal axes depict the assignment accuracies obtained and the $\Delta$ values used in the experiments, respectively. The accuracies associated with a $\Delta$ value were obtained from the classification models, each of which was created for a distinct month in the period of interest by using the same $\Delta$ value. Furthermore, the polynomials in the figures are the second degree polynomials fitted to the data. 

Looking at Figure~\ref{plot:slidingWindow}, we first observed that using issue reports from recent past to train classification models, rather than the ones from distant past, provided better assignment accuracies; the accuracies tended to decrease as $\Delta$ increased. For example, while the average assignment accuracy obtained when ${\Delta=1}$, i.e., when the issue reports submitted in the immediate preceding months were used as the training sets, was $0.73$, that obtained when ${\Delta=12}$, i.e., when the issue reports submitted in Jan $2017$ were used as the training set for the issue reports submitted in Jan $2018$, was $0.52$\%.

Looking at Figure~\ref{plot:cumulativeWindow}, we then observed that as we went back in time to collect the training data starting from the immediate preceding months (i.e., as $\Delta$ increased in the cumulative window approach), the assignment accuracies tended to increase  first and then stabilized around a year of training data. For example, while the average accuracy obtained when ${\Delta=1}$, i.e., when the issue reports submitted only in the immediate preceding months were used as the training sets, was $0.73$\%, that obtained when ${\Delta=12}$, i.e., when all the issue reports submitted in the preceding $12$ months were used as the training data set, was $0.82$\%.

Based on the results of these studies, to train a prediction model at a given point in time, we decided to use all the issue reports that have been submitted in the last $12$ months as the training set. Clearly, among all the issue reports of interest, we filter out the ones that have not yet been closed (Section~\ref{existing:setup}).

\section{Automated Issue Assignments in Practice}
\label{inaction}

In this section, we investigate our third research question (RQ3): ``How does automated issue assignment compare to manual issue  assignment in practice?'' Note that the results of this study will help evaluate the pros and cons of automated issue assignments in the field.

\subsection{Approach}
\label{inaction:approach}

To deploy IssueTAG at IsBank and Softtech, we carried out a number of meetings with the IT-HD, AST, and software development teams. In these meetings, the problems with the manual issue assignment process were discussed, IssueTAG was presented, and the effect of automating the assignment process was demonstrated by using the results of a number of preliminary studies conducted on historical data collected from the field.

One commonly accepted observation, which was made numerous times in these meetings, was that automating the issue assignment process (i.e., deploying IssueTAG) would also require to modify the other parts of the process around the deployed system to improve the efficiency and effectiveness of the entire process to the extent possible.

One refinement suggestion came from us (Process Improvement Team at Softtech). In our preliminary studies, we observed that wrong assignments made by IssueTAG were often caused due to the difficulty of distinguishing related, but different development teams from each other, such as the teams working on related products or working on different components of the same product. That is, when an issue report was assigned to a wrong team, the assignee and the correct team (i.e., the one, to which the report should have been assigned) were often related to each other, e.g., they were aware of each other's works. Consequently, we suggested that in the presence of an incorrect assignment, rather than returning the issue report to IT-HD for reassignment, which was typically the case in the manual assignment process (Section~\ref{caseDescription}), letting the assignee (e.g., the AST member embedded in the incorrectly assigned team) do the reassignment, could profoundly speed up the process. 

Another refinement suggestion came from the IT-HD management. They simply suggested to prevent IT-HD clerks from modifying the issue assignments made by IssueTAG. On one hand, this was a natural consequence of the design decision discussed above in the sense that when the reassignments are made by the current assignee, IT-HD clerks will not necessarily be aware of these modifications, thus may not learn from them to improve their assignment accuracies. On another hand, we observed that IT-HD was actually looking forward to deferring the responsibility of issue assignments. One reason was that, especially for the new IT-HD clerks, the learning curve for excelling in assignments was generally steep due to the large number of issue reports received on a daily basis and the relatively large number of development teams present (Section~\ref{deployment:analysis}). In fact, IT-HD was maintaining a knowledge base (comprised mostly of spreadsheets) to help the clerks with the assignments. However, it was cumbersome and costly for them to keep this knowledge base up to date. Nevertheless, incorrect assignments were often causing friction between the IT-HD clerks and AST members as well as the development teams. 

\subsection{Evaluation}
\label{deployment:eval}

\begin{figure}
  \includegraphics[width=1\textwidth]{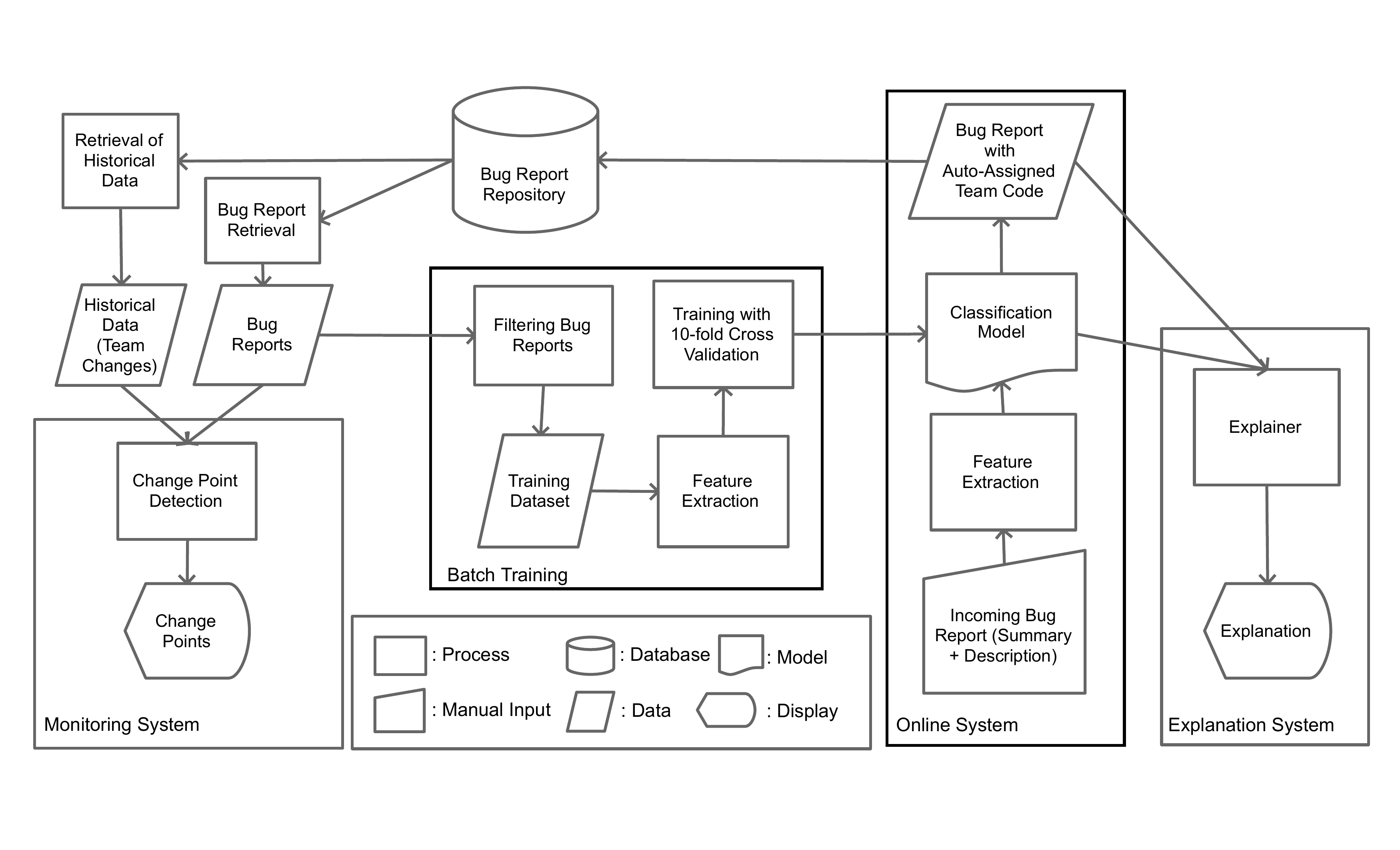}
\caption{High level architecture of IssueTAG.}
\label{proposedMLsystem}
\end{figure}

We deployed IssueTAG on Jan $12$, $2018$. The system has been fully operational since then, making automated assignments for all the issue reports submitted. Figure \ref{proposedMLsystem} presents the overall system architecture. Furthermore, Table~\ref{tbl:stats} reports some summary statistics regarding the operations of the deployed system.

\begin{table}
\centering
\caption{Summary statistics regarding the operations of IssueTAG, starting from its deployment on Jan $12$, $2018$ till June $30$, $2019$.}
\label{tbl:stats}     
\begin{tabular}{lr}
\hline\noalign{\smallskip}
Item & Value \\
\noalign{\smallskip}\hline\noalign{\smallskip}
Total number of issue reports assigned & 134,622 \\
Average number of issue reports per day & 380 \\
Total number of distinct teams & 62 \\
Average time it takes to train the model (with one year of data) & 7m 42s\\
Average response time of the system & 746 msec \\
Size of the trained model (trained with one year of data) & 588 MB \\
\noalign{\smallskip}\hline
\end{tabular}
\end{table}

\subsubsection{Deployment Setup}
\label{deployment:setup}


Based on the results of our empirical studies in Section~\ref{existing}, IssueTAG was configured to use Linear SVC to train the  classification models. And, based on the results obtained in Section~\ref{timeLocality}, the models have been trained by using the issue reports submitted in the last $12$-month time frame. Furthermore, as all the process improvement suggestions discussed in Section~\ref{inaction:approach} were accepted by all the stakeholders involved, we configured IssueTAG such that once an issue report was created by an IT-HD clerk for the first time, the report was automatically assigned to a development team by the deployed system and the IT-HD clerk did not have any means of interfering with the assignment process and/or modifying the assignment.

The system is deployed on a Dual-Core Intel(R) Xeon(R) E5-2695 v4 2.10 GHz machine with 32 GB of RAM running Windows Server 2012 R2 as the operating system.

\subsubsection{Evaluation Framework}
\label{deployment:metrics}

To evaluate the quality of the assignments over a period of time, we compute the assignment accuracy on a daily basis, which we refer to as {\em daily assignment accuracy}. More specifically, the daily assignment accuracy achieved on a day $d$, is the ratio of the assignments that are correctly made for the issue reports opened on the day $d$. Note that we compute the daily accuracies based on the dates, on which the issue reports are opened, rather than they are closed. This is because the automated assignments are made as soon as the issue reports are created (i.e., opened) by using the underlying classification model, which was available at the time of the creation.

To evaluate the reduction in the amount of manual effort required for the issue assignments, we measure the person-months saved by automating the process. To this end, a survey we conducted on the IT-HD clerks revealed that, given an issue report, it takes about $30$ seconds on average for an IT-HD clerk to assign the report to a development team, which is mostly spent for reasoning about the issue report and (if needed) performing a keyword-based search in the knowledge base. Note that this effort does not include the effort needed to maintain the knowledge base. Therefore, the actual amortized manual effort is expected to be higher than $30$ seconds. IssueTAG, on the other hand, requires no human intervention to make an assignment once an issue report has been created.

To evaluate the effect of the deployed system as well as the improvements made in the issue assignment process, we compute and compare the {\em solution times} before and after the deployment of IssueTAG. In particular, we define the solution time for an issue report as the time passed between the report is opened and it is closed. The shorter the solution times, the better the proposed approach is. Furthermore, as the characteristics of the reported issues, thus the solution times, can change over time, we, in the evaluations, compute and compare the solution times for the issue reports that were opened within two months before and after the deployment of IssueTAG.


\subsubsection{Data and Analysis}
\label{deployment:analysis}

\begin{figure}[t]
\centering
\includegraphics[width=1\textwidth]{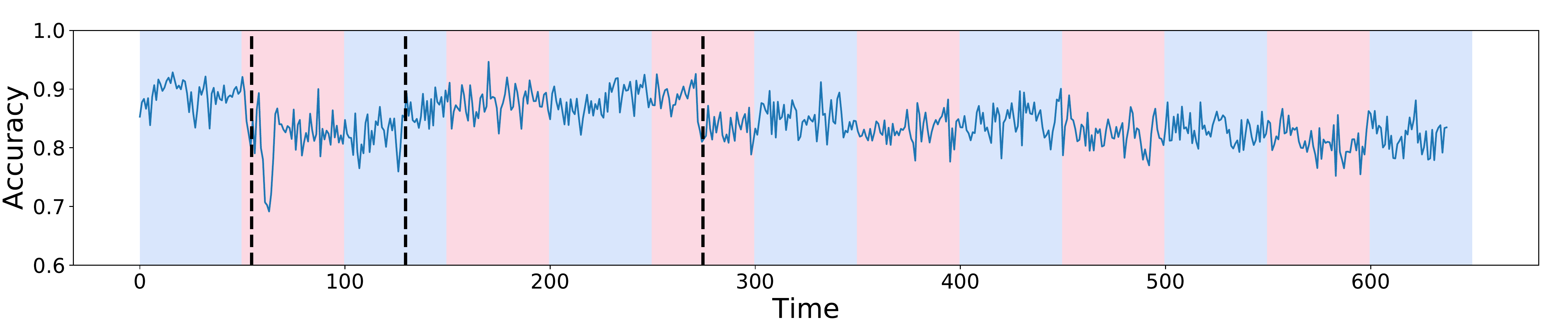}
\caption{Daily assignment accuracies achieved between December 2016 and June 2019. The time point $0$ represents the date, on which the manual issue assignment process as it is described in Section~\ref{caseDescription} was started. The vertical dashed lines represent the points in time where a shift in daily accuracies was automatically detected by our change point detection approach (Section~\ref{monitor}). IssueTAG was deployed at the third dashed line, i.e., all the accuracies before this line were obtained by manual assignments, whereas those after were obtained by auomated assignments. The first dashed line represents the date, on which significant changes in team responsibilities occurred due to migrating certain functionalities from mainframes to state-of-the-art platforms. Therefore, the time gap between the first and second dashed lines (i.e., about $2.5$ months) represent the amount of time it took for the IT-HD clerks to adapt to these changes.}
\label{deployment:accuracies}
\end{figure}

Figure~\ref{deployment:accuracies} presents the daily assignment accuracies achieved between December 2016 and June 2019. The time point $0$ in this figure represents the date, on which the manual issue assignment process as it is described in Section~\ref{caseDescription}, was started. Furthermore, the vertical dashed lines in the figure represent the points in time where a shift in daily accuracies was automatically detected by the change point detection approach we had developed (Section~\ref{monitor}). IssueTAG was, indeed, deployed exactly at the $273$th time point where the third vertical dashed line resides. That is, all the accuracies before this dashed line were obtained by manual assignments, whereas those after were obtained by automatic assignments. The other vertical dashed lines will be discussed below in this section.



We first observed that after IssueTAG was deployed, the daily assignment accuracies dropped slightly (Figure~\ref{deployment:accuracies}). More specifically, the average daily accuracies before and after the deployment were $0.864$ (${min=0.691}$, ${max=0.947}$, ${stddev=0.040}$) and $0.831$ (${min=0.752}$, ${max=0.912}$, \linebreak ${stddev=0.027}$), respectively.

We, however, observed that the accuracy of an automated issue assignment system does not have to be higher than that of manual assignments in order for the system to be useful. First, we observed that IssueTAG reduced the manual effort required for the assignments. In particular, given that it takes an average of $30$ seconds for an IT-HD clerk to assign an issue report to a development team and an average of $8,000$ issue reports are received on a monthly basis, IssueTAG has been saving $5$ person-months yearly, on average ($8,000$ issue reports * $30$ seconds = $240,000$ seconds per month = $5$ person-months per year).






Second, we observed that the deployed system together with the process improvements we implemented, profoundly reduced the turnaround time for closing the issue reports. More specifically, the average solution times before and after the deployment were $3.26$ days and $2.61$ days, respectively.


Third, we observed that it can take quite a while for a human stakeholder to excel in the issue assignment task, which is, in deed, a problem, especially in the presence of high employee turn over rates. For example, the first vertical dashed line in Figure~\ref{deployment:accuracies}, represents the date on which an integral part of the core banking system was migrated from mainframes to state-of-the-art hardware and software platforms. As a result of this migration, the structure and the responsibilities of the related development teams changed significantly. In particular, the responsibilities of one development team working on mainframes were migrated to $3$ development teams working on state-of-the-art platforms, which consisted of completely different software engineers. Evidently, the assignment accuracies were affected by this change; the daily accuracies dropped at the first vertical dashed line (i.e., $55$th time point) and stayed low until the second vertical dashed line (i.e., the $130$th time point). More specifically, the average daily accuracies obtained from the manual assignments before the first dashed line, in between the first and  second dashed lines, and after the second dashed line until IssueTAG was deployed at the third dashed line were, $0.889$ (${min=0.825}$, ${max=0.929}$, ${stddev=0.024}$), $0.819$ (${min=0.691}$, ${max=0.900}$, ${stddev=0.039}$), and $0.879$ (${min=0.822}$, ${max=0.947}$, ${stddev=0.024}$), respectively. That is, it took the IT-HD clerks about $2.5$ months to adapt to the new development teams. Therefore, this time frame can be considered to be a lower bound on the amount of time a new hire would require to learn to make accurate assignments. It is a lower bound in the sense that only $19$\% of the issue reports were affected by the changes in the team responsibilities during the aforementioned period of time and that the IT-HD clerks already had a great deal of experience; for a new hire, everything will be new.

Note further that the $0$th time point in Figure~\ref{deployment:accuracies} represents the date, on which Jira was started to be used for storing and managing the issue reports. That is, IT-HD clerks had been making manual assignments before this date, but had different means of managing the reports, which explains the high daily assignment accuracies event at the $0$th time point in the figure. As we didn't have any access to the issue databases maintained before the $0$th time point, we used only the issue reports managed by Jira in this research.


\section{User Evaluations}
\label{userEvals}

In this section, we investigate our fourth research question (RQ4): ``Is IssueTAG perceived as useful by the end-users?''

\subsection{Approach}
\label{userEvals:approach}

To carry out the study, we created a survey by following a survey template frequently used at Softtech. It had a total of $8$ questions from two categories: {\em requirement satisfaction} and {\em product quality}. The former category aims to evaluate the extent to which the deployed system meets its requirements, whereas the latter category aims to evaluate the quality of the final product. All questions, except for the last one, were Likert scale questions each with answer options: {\em no opinion}, {\em 1 - strongly disagree}, {\em 2 - disagree}, {\em 3 - agree}, and {\em 4 - strongly agree}. The last question was an open-ended question. Furthermore, for the Likert scale questions, we asked the participants to elaborate on their responses, if they had ``disagreed'' or ``strongly disagreed.'' Table~\ref{tbl:generalSurvey} presents the questions we used in the survey.


\begin{table}
\caption{Survey questions used for evaluating IssueTAG.}
\label{tbl:generalSurvey}
\resizebox{\columnwidth}{!}   
{\begin{tabular}{lllr}
\hline\noalign{\smallskip}
No & Question & Type & Category \\
\noalign{\smallskip}\hline\noalign{\smallskip}
Q1 & I know the business requirements that & &  \\
 & the system is supposed to meet. & Likert scale & requirements satisfaction \\
Q2 & The system (as a software product) is reliable. & Likert scale & requirements satisfaction \\
Q3 & The system is useful. & Likert scale & requirements satisfaction \\
Q4 & The system reduces the solution times for issue reports. & Likert scale & product quality  \\
Q5 & The issue assignments made by the system are trustworthy. & Likert scale & product quality \\
Q6 & The system is robust. & Likert scale & product quality \\
Q7 & I recommend the system to other companies. & Likert scale & product quality \\
Q8 & What do you like and don't like about the system? & &\\
   & Do you have any suggestions for improvement? & open-ended & product quality \\
\noalign{\smallskip}\hline
\end{tabular}}
\end{table}

\subsection{Evaluation}
\label{userEvals:evals}

We conducted the survey on the AST members. We chose this group of stakeholders as the recipients of the survey because, being embedded in the development teams, they were the direct end-users of IssueTAG. That is, they, as the first recipients of the issue reports, were the ones to validate whether the assignments were correct or not and to reassign them as needed. The IT-HD clerks, on the other hand, could not participate in the survey because they were not considered to be the end-users of the deployed system in the sense that they neither made use of the assignments automatically made by the deployed system nor had a control over them.

\subsubsection{Experimental Setup}
\label{userEvals:setup}

About half of the AST members (more specifically, $14$ out of $30$) voluntarily agreed to participate in the study. The participants filled out the survey online at their spare time.

\subsubsection{Evaluation Framework}
\label{userEvals:metrics}

For the Likert scale questions, we use the frequencies and the average scores obtained to quantitatively analyze the results. The average scores were computed as the arithmetic average of the scores with the ``no opinion'' responses excluded. For the open-ended question, we present the answers we received (Table~\ref{tbl:openEndedQuestion}) and qualitatively discuss them.

\subsubsection{Data and Analysis}
\label{userEvals:analysis}

\begin{figure*}
\includegraphics[width=0.96\textwidth]{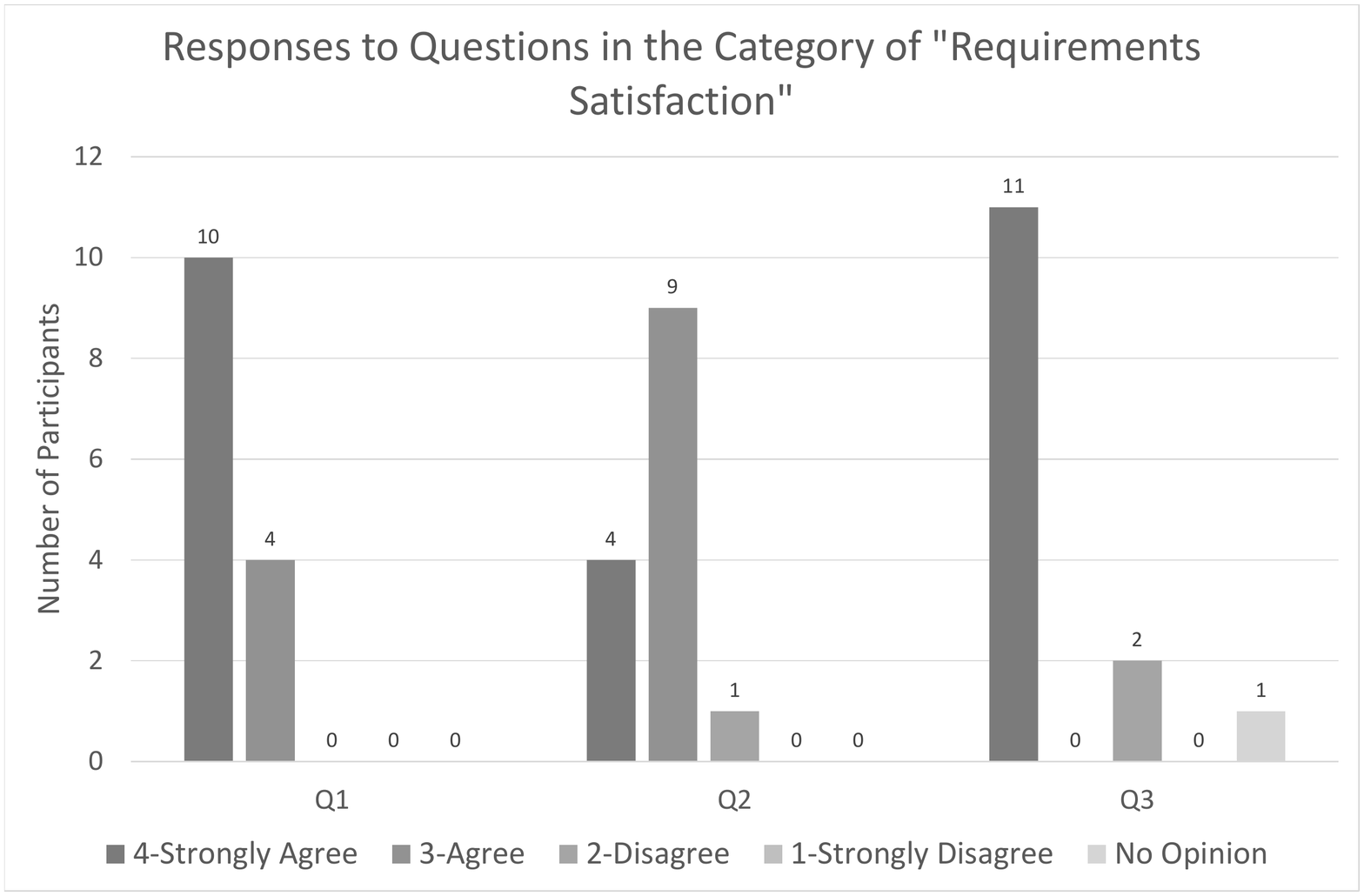}
\caption{Responses to questions in the category of ``requirements satisfaction.''}
\label{fig:requirementsSatisfaction}       
\end{figure*}

The results of the survey strongly suggest that IssueTAG meets its business needs with high quality. Regarding the questions in the category of ``requirements satisfaction,'' we observed that the majority of the participants thought IssueTAG was useful and reliable (Figure~\ref{fig:requirementsSatisfaction}). More specifically, all of the participants ``strongly agreed'' or ``agreed'' to Q1, indicating that they knew the business requirements that IssueTAG was supposed to meet. And, $92.86$\% ($13$ out of $14$) of the participants for Q2 and $78.57$\% ($11$ out of $14$) of the participants for Q3, responded ``agree'' or higher. The average scores were $3.71$, $3.21$, and $3.69$ (out of $4$) for these questions, respectively.

\begin{figure*}
\includegraphics[width=0.96\textwidth]{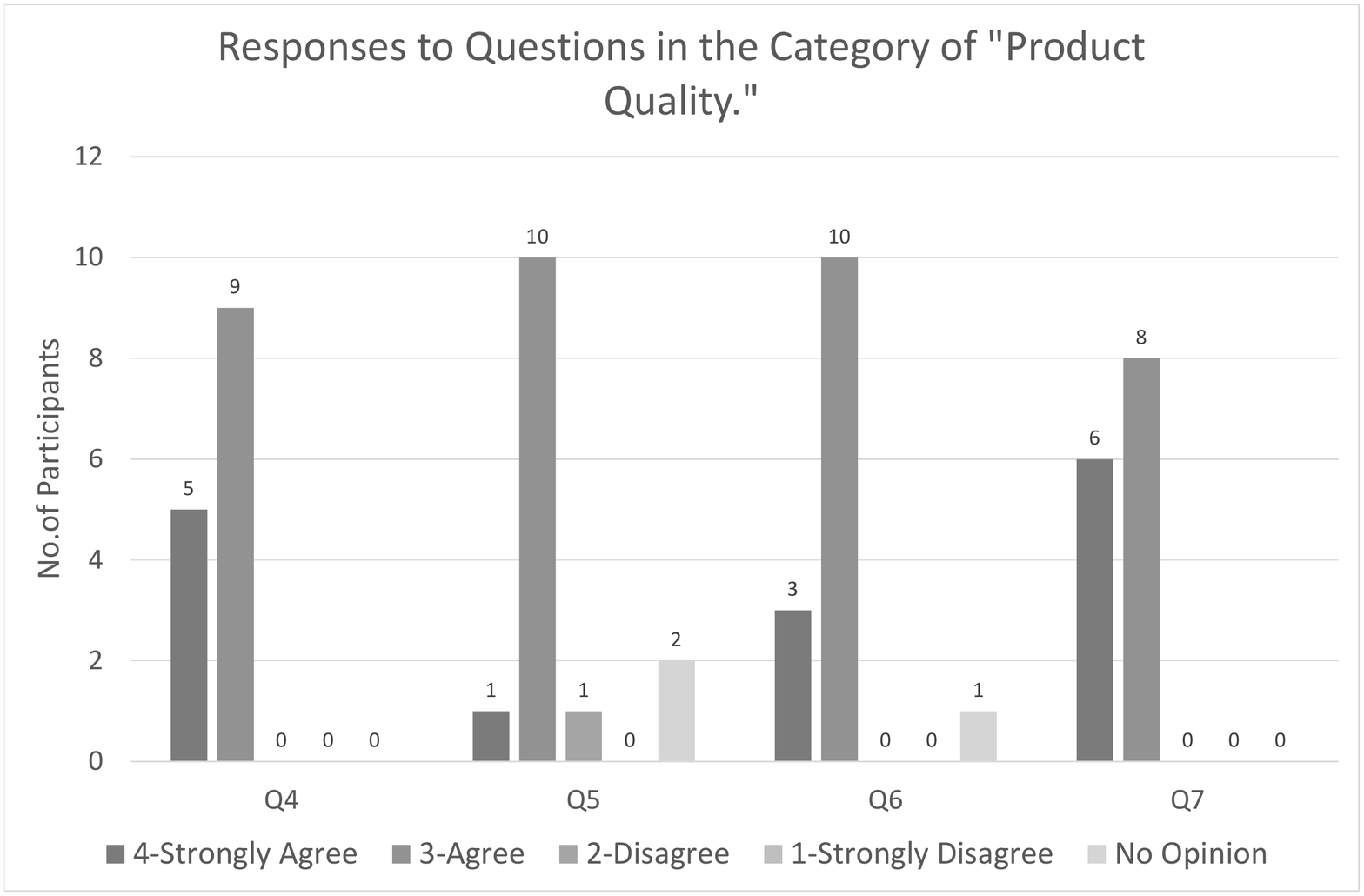}
\caption{Responses to questions in the category of ``product quality.''}
\label{fig:productQuality}     
\end{figure*}

Only $1$ participant for Q2 and $2$ participants for Q3 ``disagreed.'' The comments that they provided as to why they disagreed are given in Table~\ref{tbl:disagreements}. Evidently, part of the reason was that these participants were unrealistically expecting to have perfect assignments (with no incorrect assignments) from the deployed system.

\begin{table}
\caption{Comments that the participants provided as to why they ``disagreed.''}
\label{tbl:disagreements}
\resizebox{\columnwidth}{!}   
{\begin{tabular}{rrl}
\hline\noalign{\smallskip}
Question & Comment & Comment \\
No & No & \\
\noalign{\smallskip}\hline\noalign{\smallskip}
Q2 & 1 & Due to some keywords [appearing in issue reports], the system \\
 & & sometimes make incorrect assignments to my team.\\
Q3 & 1 & In fact, the application is useful. However, in the presence of a \\ 
 & & wrong assignment made by the system, reassigning the bug \\
  & & report to the correct team, especially when we don't know \\
 & &  which team it should really be assigned to or when the other \\ 
 & & team refuses to take the responsibility for the issue report, \\
 & & causes delays in the process.\\
Q3 & 2 & The system sometimes makes incorrect assignments.\\
Q5 & 1 & I can't say ``I agree'' because I sometimes encounter incorrect \\
  & & assignments.\\
\noalign{\smallskip}\hline
\end{tabular}}
\end{table}

Regarding the other quality aspects of the system, $100$\% ($14$ out of $14$) of the participants for Q4, $78.57$\% ($11$ out of $14$) for Q5, $92.86$\% ($13$ out of $14$) for Q6, and $100$\% ($14$ out of $14$) for Q7 responded ``agree'' or higher (Figure~\ref{fig:productQuality}). The average scores were $3.36$, $3.0$, and $3.0$, and $3.43$ (out of $4$) for these questions, respectively. Only $1$ participant disagreed with Q5, the comment of whose can be found in Table~\ref{tbl:disagreements}. We, furthermore, observed that all the participants would recommend the system to other companies; all responded ``agree'' or higher to Q7 (Figure~\ref{fig:productQuality}).

\begin{table}
\caption{Responses given to the open-ended question Q8.}
\label{tbl:openEndedQuestion}
\resizebox{\columnwidth}{!}   
{\begin{tabular}{rl}
\hline\noalign{\smallskip}
Comment & Comment \\
No & \\
\noalign{\smallskip}\hline\noalign{\smallskip}
1 & I think that the assignments are made rapidly and accurately. I have not been having any issues with the \\
& system. We [as a team] rarely receive incorrect assignments. I, however, believe that this is normal because \\
& the same words [terms] can be related with multiple development teams. It is quite normal for the system \\
& not being able to distinguish between teams in such situations.\\

2 & Most of the time, the system works for us. Sometimes, however, it assigns irrelevant issue reports to my team.\\

3 & General words, such as ``problem'', should not be used by the system when assigning issue reports to \\
& development teams.\\

4 & The system profoundly reduced the loss of time by rapidly assigning issue reports to development teams \\
& with high accuracy.\\

5 & I think that it takes a while for an AI algorithm to learn about new issue reports. Improvements can be made \\ 
& in this area.\\

6 & I believe that the system had a profound effect on assigning the issue reports to the right development teams.\\

7 & It is a nice and practical system, better results can be obtained with further development. One area for \\
& improvement could be to explain the keywords [terms] used for the assignments.\\
\noalign{\smallskip}\hline
\end{tabular}}
\end{table}

Last but not least, the responses given to the open-ended question Q8 can be found in Table~\ref{tbl:openEndedQuestion}. All of these comments can be considered as generally positive. A couple of them actually make some suggestions for future improvements. For example, the last comment basically suggests that the system should provide an explanation as to why a given issue report is assigned to the selected development team. As a matter of fact, this request turned out to be a common one, for which we have developed an automated approach (Section~\ref{explain:approach}).




 
\section{Explaining Team Assignments}
\label{explain}

One interesting observation we made after IssueTAG had been deployed was that, occasionally, especially for incorrect assignments, the stakeholders demanded some explanations as to why and how certain issue reports had been assigned to their teams. This was an issue we didn't expect to face before deploying the system. As a matter of fact, based on the informal discussions we had with the stakeholders, we quickly realized that explaining the assignments could further improve the trust in IssueTAG. 

In this section, we develop and empirically evaluate (by conducting a survey on actual stakeholders) an approach for automatically generating explanations for the issue assignments made by the underlying classification model, answering our fifth research question (RQ5): ``Can the issue assignments made by the underlying data mining model be explained in a non-technical manner?''

Note that since the classification models we use, namely the linear SVC models, are not human-readable, providing such explanations is a non-trivial task. To the best of our knowledge, there is, indeed, no work in the literature of automated issue assignment, addressing this problem. 

One requirement we have is that the explanations should easily be interpreted and understood even by non-technical stakeholders as the recipients of these explanations are not necessarily technical stakeholders. Another requirement is that they should be given in terms of the natural language descriptions present in the issue reports, so that stakeholders can relate to them.

With all these in mind, we conjecture that providing a list of most influential (positive or negative) words for an issue assignment together with their relative impact scores as an explanation for the assignment, could help stakeholders understand the rationale behind the assignments. 

Interestingly enough, we observe that such explanations could also be used in an interactive manner to enable the stakeholder creating the issue report to provide feedback to the classification model. Although such human-in-the-loop assignments are out of the scope of the this paper, we, nevertheless, added additional questions to our survey to evaluate the plausibility of the idea.






\subsection{Approach}
\label{explain:approach}

We use LIME (Local Interpretable Model-Agnostic Explanations) to automatically produce explanations for the issue assignments made by IssueTAG. LIME is a model-agnostic algorithm for explaining the predictions of a classification or regression model \citep{ribeiro2016should}. In this work, we, (to the best of our knowledge) for the first time, use LIME in the context of automated issue assignment and evaluate it by carrying out a survey on actual stakeholders in the field. Next, we briefly describe the LIME algorithm without any intention to provide all the mathematics behind it. The interested reader can refer to~\citep{ribeiro2016should} for further details.

LIME, in our context, aims to identify a human-interpretable, locally faithful model, which provides qualitative understanding between the terms used in issue reports and the development teams, to which they are assigned. In a nutshell, given an issue report, the assignment made for this report, and the underlying classification model, LIME first represents the report as a bag of words and samples instances around the report by drawing subsets of the words in the bag uniformly at random. Then, the samples are weighted by their proximities to the original issue report and fed to the classification model to label them. Next, all the samples together with their associated labels are used to learn a linear model comprised of $K$ terms (in our case, ${K=6}$), which distinguishes the labels. Finally, the linear model learnt is reported as an explanation for the assignment.


The explanation generated for an assignment is, indeed, a set of $K$ terms selected from the original issue report together with their relative weights, indicating the influential terms that either contribute to the assignment or are evidence against it. Figure~\ref{fig:goodExplanation}a presents an example explanation created for an assignment made by IssueTAG in the field. The vertical axis reports the most influential terms selected, whereas the horizontal axis denotes their relative weights. The terms with positive weights depict the terms that contribute to the assignment, wheres as those with negative weights depict the ones that are evidence against it. That is, in a sense, the former set of terms vote for the assignment, whereas the latter ones vote against it in an attempt to change the assignment.

\subsection{Evaluation}
\label{explain:evaluations}

We could not simply ask the stakeholders to evaluate each and every explanation created for the issue assignments, which were of interest to them. The reason was that there was a large number of issue reports submitted on a daily basis  (Section~\ref{deployment:eval}) and that checking out the explanations was optional, i.e., the stakeholders were not required to have a look at the explanations. Therefore, forcing the stakeholders to evaluate the explanations as the issue assignments were made, could have adversely affected their performance.
 
\subsubsection{Experimental Setup}
\label{explain:setup}
  
We, therefore, carried out an empirical study with the same participants involved in our survey in Section~\ref{userEvals}, after having their consensus to voluntarily participate in this additional study, which were accepted by all of them. 

For each participant, we randomly picked $10$ issue assignments, which were handled by the participant in the last week before the study, such that the ratio of correctly and incorrectly assigned issue reports roughly resembled the average daily assignment accuracy. When there were less than $10$ issue assignments for a participant, we selected all of the available ones. All told, we picked a total of $130$ issue assignments ($10$ for each participant, except for two, for whom we could have only $5$ assignments each). Out of all the selected assignments, $13$ ($10$\%) were incorrect. 


We then created a questionary for each participant by using the issue assignments selected for the participant. For each assignment in the questionary, we included 1) the issue report, 2) the assignment made by IssueTAG, 3) the explanation automatically created by the proposed approach, using the $6$ most influential terms involved in the assignment, and 4) four questions (Table~\ref{tbl:explanationSurvey}).

\begin{table}
\caption{Survey questions related to selected issue reports and their explanations.}
\label{tbl:explanationSurvey}
\resizebox{\columnwidth}{!}   
{\begin{tabular}{lll}
\hline\noalign{\smallskip}
No & Question & Type\\
\noalign{\smallskip}\hline\noalign{\smallskip}

Q1 & Is the explanation helpful in understanding the assignment? & Yes/No \\

Q2 & Given the issue report, the assignment, and the explanation for the assignment, & \\
& how would you rate the trustworthiness of the assignment? & Likert scale \\

Q3 & Which terms in the explanation make you think that the assignment is not trustworthy? & Open ended \\

Q4 & What are the, additional terms that you would like to see in the explanation; & \\
&  or the terms, the impact factor of which you would like to increase in the explanation, & \\
& before you can trust the assignment? & Open ended \\

\noalign{\smallskip}\hline
\end{tabular}}
\end{table}

The first two questions, namely Q1 and Q2, were directly concerned with our main research question in this study, i.e., whether or not the automatically generated explanations could help stakeholders understand the rationale behind the assignments. Q1 was a ``yes'' or ``no'' question, whereas Q2 was a Likert scale question with answer options: {\em 2 - very trustworthy}, {\em 1 - trustworthy}, {\em 0 - not sure}, {\em -1 - untrustworthy}, {\em -2 - very untrustworthy}. The last two questions, namely Q3 and Q4, on the other hand, aimed to evaluate the plausibility of using the explanations to get feedback from the stakeholders in an attempt to further improve the assignment accuracies. These questions were open-ended questions, which were conditional on Q2; the participants were asked to answer these questions only when the response to Q2  was either ``untrustworthy'' or ``very untrustworthy.''


All the explanations were created by using the {\tt LIME} Python library~\citep{ribeiro2016should} with ${K=6}$  --  a decision we made based on the maximum number of terms that we thought a stakeholder could efficiently and effectively reason about.

\subsubsection{Evaluation Framework}
\label{explain:eval}

For Q1 and Q2, we use the frequencies of responses to quantitatively analyze the results. For Q3 and Q4 (when answered), we manually investigate how the feedbacks can be used to further improve the accuracies.


\begin{figure*}
\includegraphics[width=0.96\textwidth]{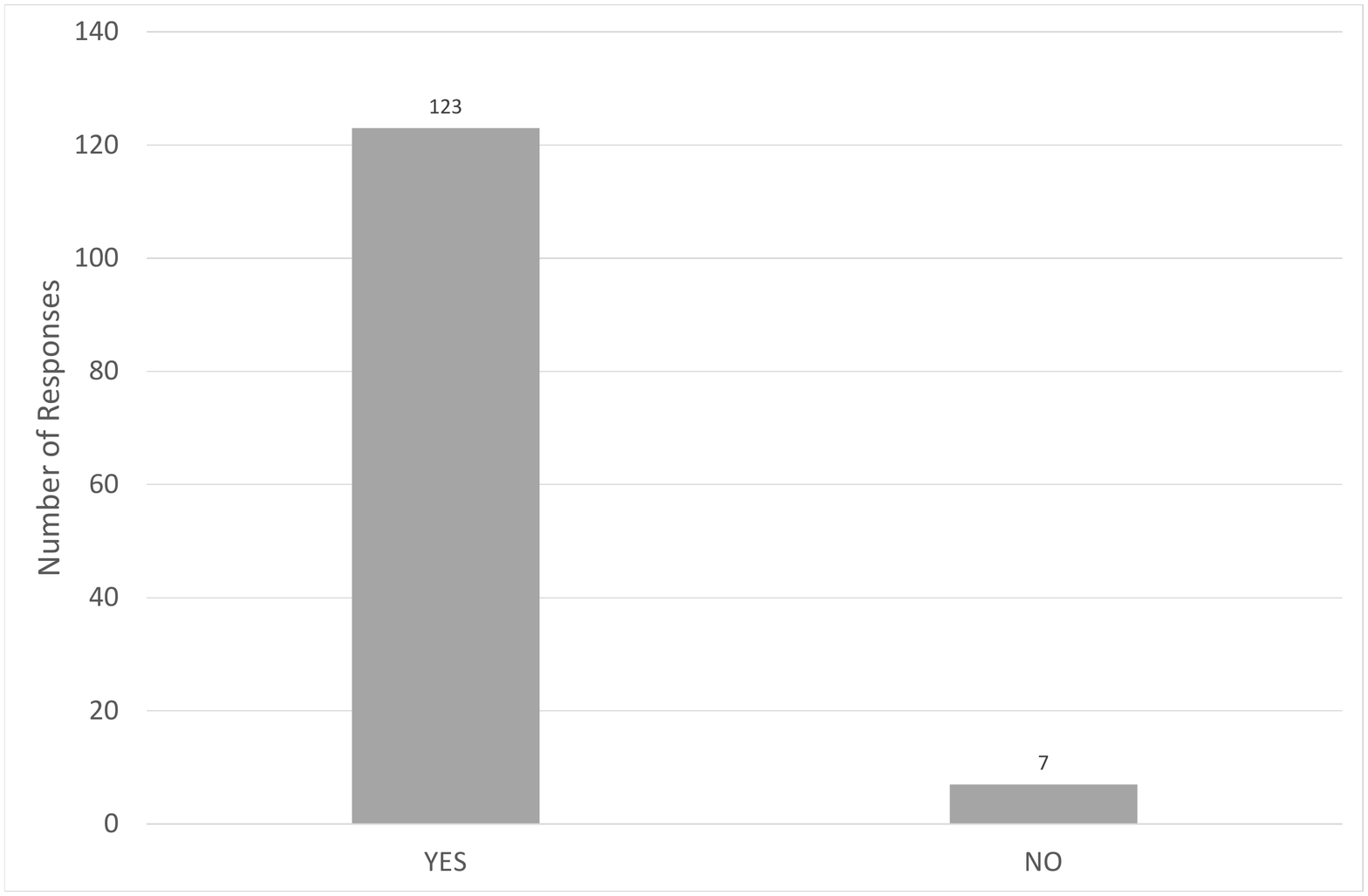}
\caption{Responses to Q1: ``Is the explanation helpful in understanding the assignment?''}
\label{fig:explanationQ1}       
\end{figure*}

\begin{figure*}
\includegraphics[width=0.96\textwidth]{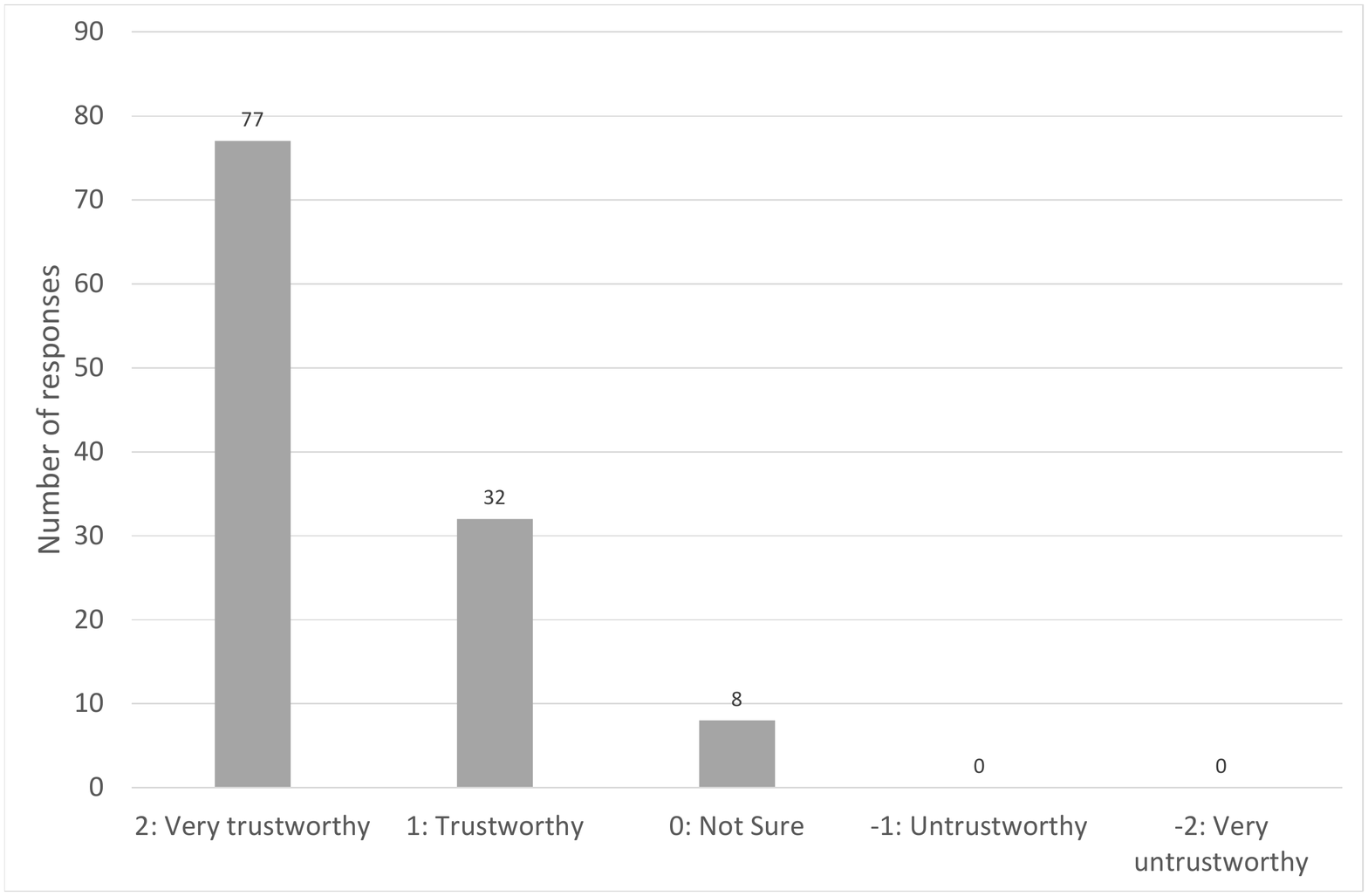}
\caption{Responses to Q2 (for the correct assignments): ``Given the issue report, the assignment, and the explanation for the assignment, how would you rate the trustworthiness of the assignment?''}
\label{fig:explanationQ2a}       
\end{figure*}

\begin{figure*}
\includegraphics[width=0.96\textwidth]{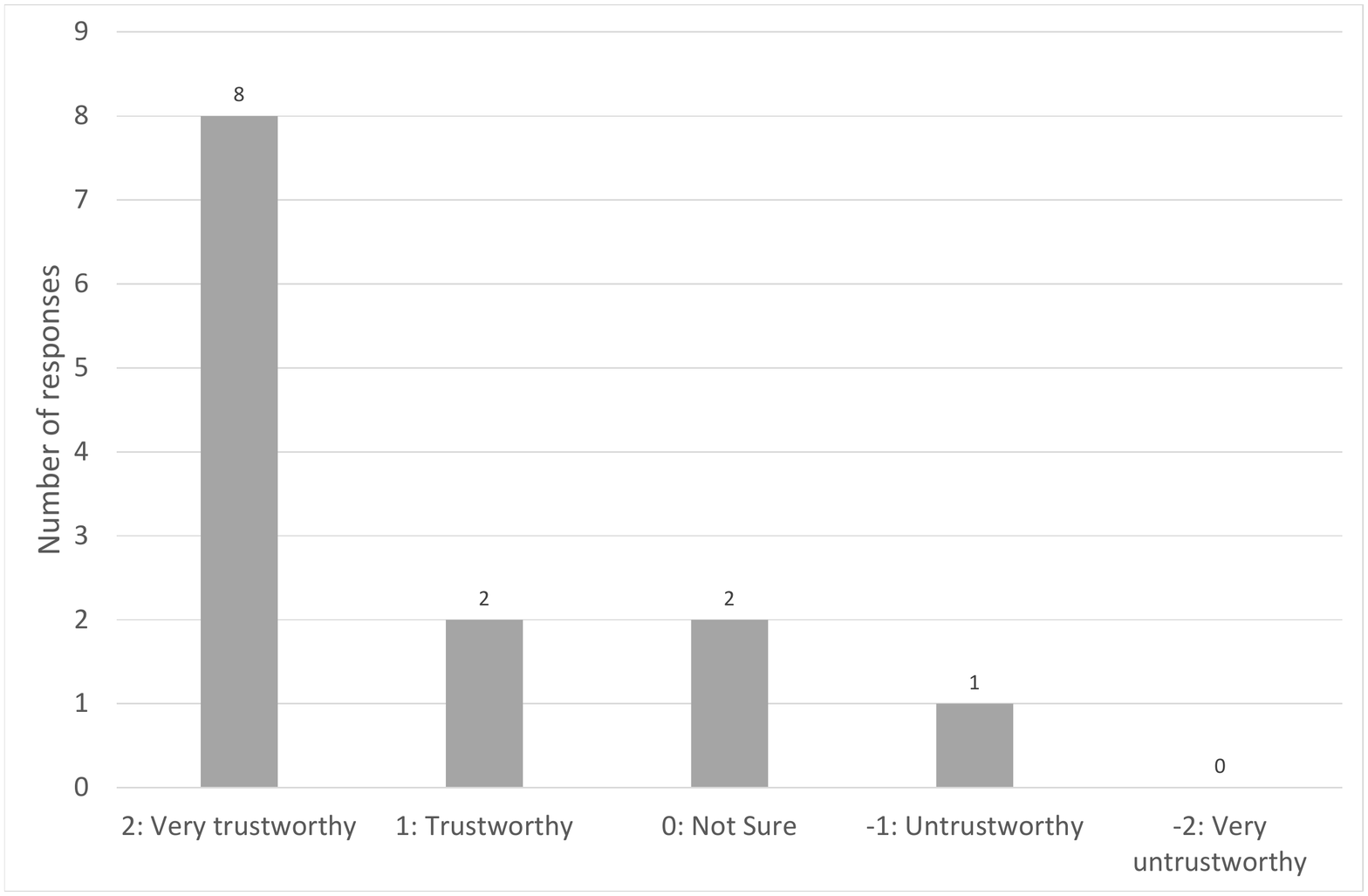}
\caption{Responses to Q2 (for the incorrect assignments): ``Given the issue report, the assignment, and the explanation for the assignment, how would you rate the trustworthiness of the assignment?''}
\label{fig:explanationQ2b}       
\end{figure*}

\subsubsection{Data and Analysis}
\label{explain:analysis}

Regarding Q1, we observed that participants found $95$\% ($123$ out of $130$) of the explanations, each of which was created for a distinct issue assignment, helpful in understanding the rationale behind the assignments (Figure~\ref{fig:explanationQ1}). 


Regarding Q2, based on the explanations created for the correct assignments, the participants found $93$\% of the assignments ($109$ out of $117$) ``trustworthy'' or ``very trustworthy'' (Figure~\ref{fig:explanationQ2a}). And, for the remaining $7$\% of the assignments ($8$ out of $117$), they were ``not sure'' whether the explanations helped them decide if the assignments were reliable or not. None of the assignments was found ``untrustworthy'' or ``very untrustworthy.''

Interestingly enough, based on the explanations created for the incorrect assignments, we observed that the participants found $77$\% of the assignments ($10$ out of $13$)  ``trustworthy'' or ``very trustworthy,'' suggesting that given the same issue reports, these participants would have made the same or similar mistakes in assigning the reports. We believe that this was because of some missing information in these issue reports, which was required for accurate assignments (Figure~\ref{fig:explanationQ2b}). Furthermore, the participants were ``not sure'' about the trustworthiness of the $15$\% of the assignments ($2$ out of $13$).

\begin{figure}[htb]
\centering
  \subfloat[]{
    \includegraphics[width=1\textwidth]{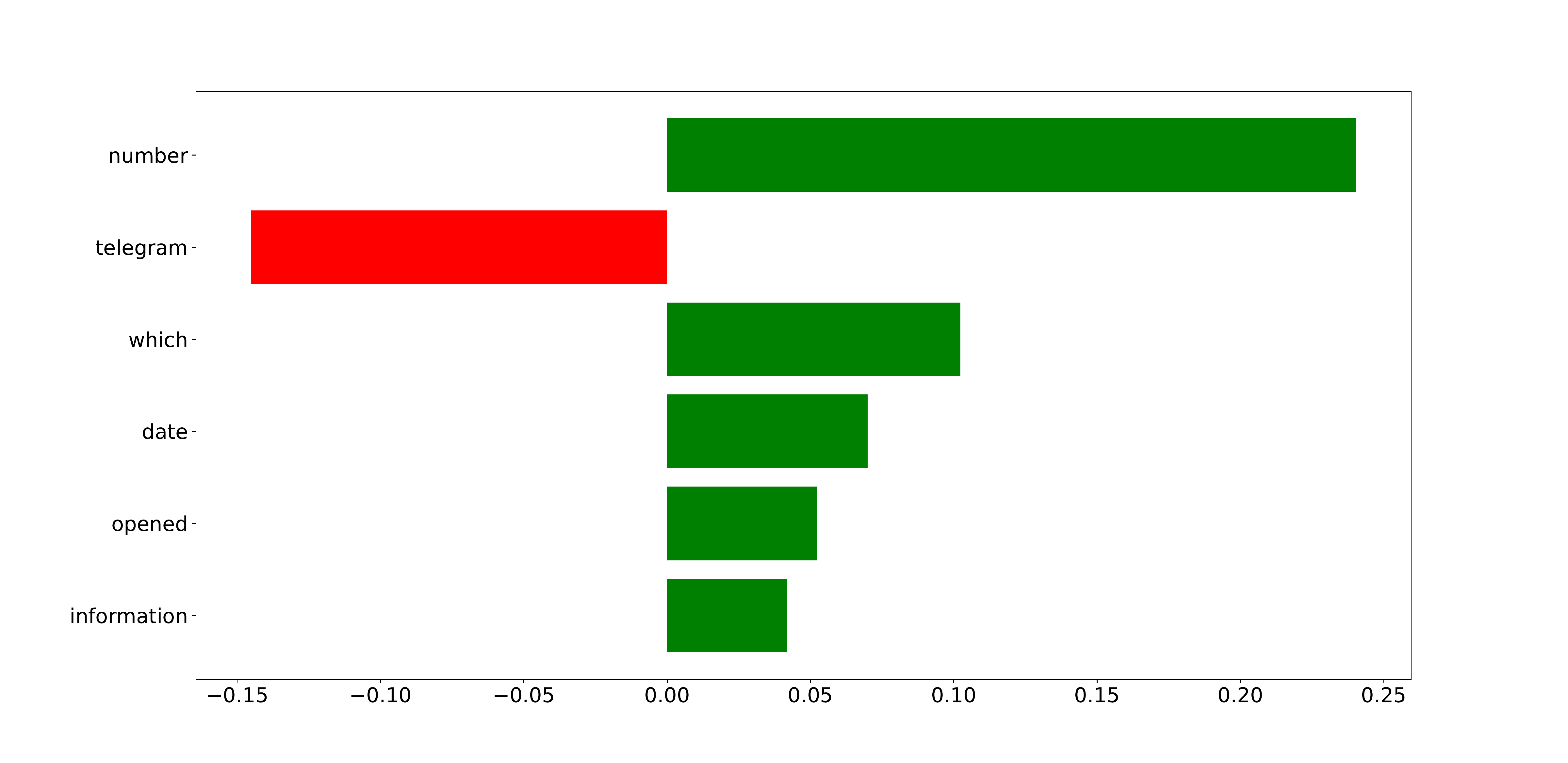}}\hfill
  \subfloat[]{
    \includegraphics[width=1\textwidth]{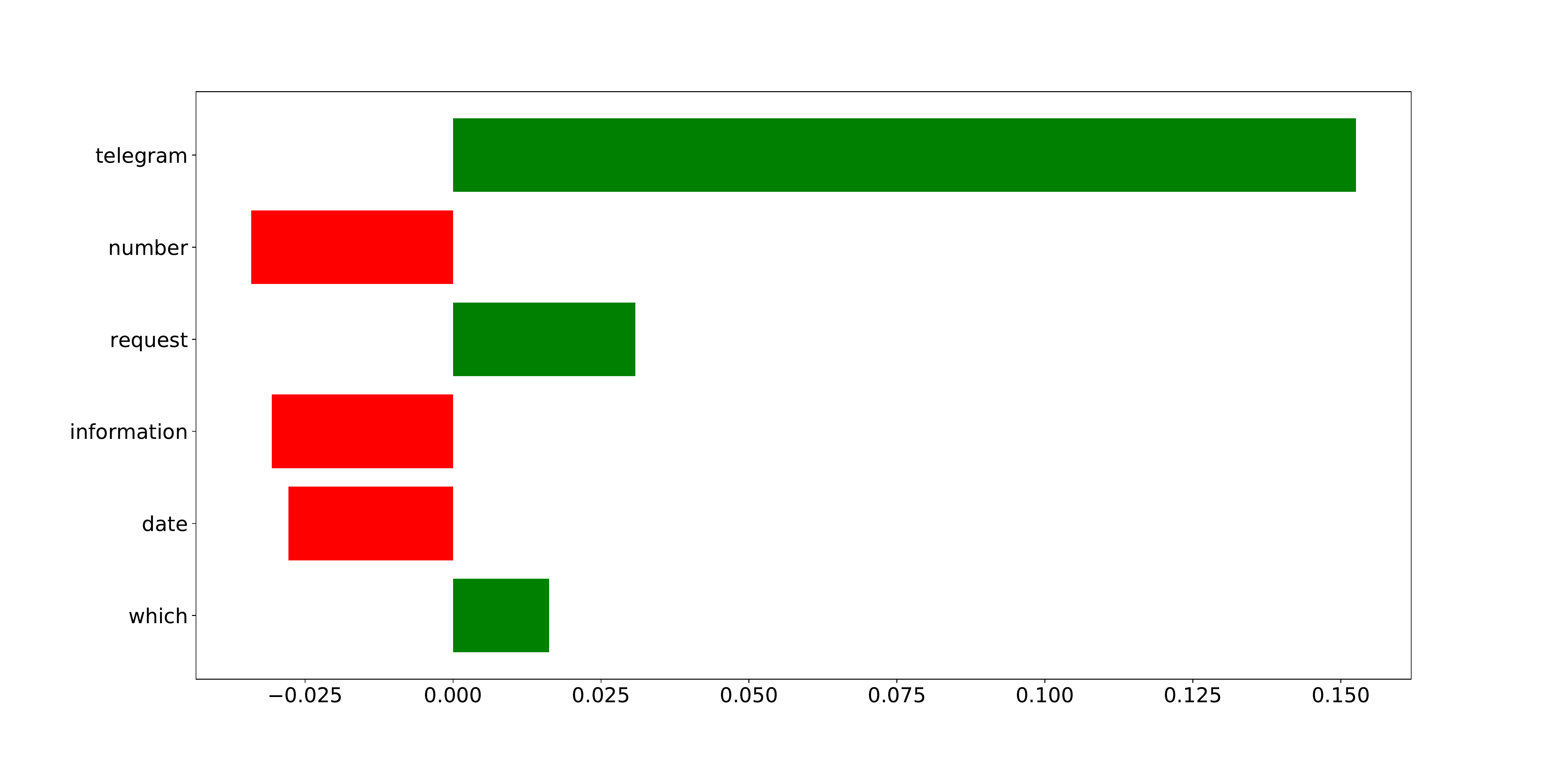}}\hfill
  \caption{The explanations created for the assignment marked as ``untrustworthy'' by a participant: a) the explanation created for the original assignment, which was incorrect and b) the explanation created for the second likely assignment, which was correct.}
\label{fig:goodExplanation}
\end{figure}

Regarding Q3 and Q4, among all the responses given to Q2, only one was scored below $0$. That is, based on the explanations created for the assignments, only one of the assignments was found ``untrustworthy.'' And, this assignment was, indeed, an incorrect assignment made by IssueTAG.

The explanation created for the aforementioned assignment is given in Figure~\ref{fig:goodExplanation}a. Given this explanation, the participant argued in her response that the term ``telegram,'' which is a domain specific term used when creating a credit account, was an important term for the issue report at question. Therefore, it should have positively, rather than negatively, affected the assignment. As a matter of fact, this argument was also well-aligned with the automatically generated explanation given in Figure~\ref{fig:goodExplanation}a in the sense that ``telegram,'' being a term with a large negative impact, voted against the assignment in an attempt to change it. It was, however, not strong enough to modify the outcome.

Interestingly enough, Figure~\ref{fig:goodExplanation}b presents the explanation created for the second likely assignment made by the underlying classification model, which turned out to be the correct assignment. Note that in this assignment, the term ``telegram'' had  the largest positive impact on selecting the correct team, which was also suggested by the stakeholder. Therefore, had the participant presented with the explanations created for the top two most likely assignments, she could have selected the second assignment, thus increased the assignment accuracy. Note that the aforementioned type of approaches are beyond the scope of this work. However, as the results of this study are promising, we, as a future work, plan to develop ``human-in-the-loop'' approaches, which leverage the automatically created explanations to further improve the assignment accuracies.


\section{Monitoring Deterioration}
\label{monitor}

In this study, we address our sixth research question (RQ6): ``Can the deteriorations in the assignment accuracies be automatically detected in an online manner?'' This was, indeed, another issue we faced after the deployment of IssueTAG. It is important because such a mechanism not only increases the confidence of the stakeholders in the system, but also helps determine when the underlying classification model needs to be recalibrated by, for example, retraining the model (Sections~\ref{existing}-\ref{timeLocality}).

\subsection{Approach}
\label{monitor:approach}

One observation we make is that every issue report at Softtech is closed by the development team, who has fixed the reported issue. Therefore, in the presence of an incorrect assignment made by IssueTAG, the report is reassigned and the history of the reassignments is stored in the issue tracking system. Consequently, at any point in time, the assignment accuracy of IssueTAG can automatically be computed using the history of the issue reports that have been closed.  Therefore, deteriorations in the accuracy can be analyzed in an online manner.


To this end, we use an online change point detection approach, called  Pruned Exact Linear Time (PELT) \citep{killick2012optimal}. In a nutshell, PELT is a statistical analysis technique to identify when the underlying model of a signal changes \citep{truong2018selective}. In our context, we feed PELT with a sequence of daily assignment accuracies (Section~\ref{monitor:eval}) as the signal. The output is a set of points in time (if any) where mean shifts. PELT, being an approach based on dynamic programming, detects both the number of change points and their locations with a linear computational cost under certain conditions \citep{killick2012optimal}. Further information can be found in~\citep{killick2012optimal} and~\citep{truong2018selective}.

PELT has been used for change point detection in many application domains, including DNA sequence data, financial time series, and oceanographic data \citep{hocking2013learning, lavielle2007adaptive, killick2012optimal}. In this work, we, on the other hand, use it (to the best of our knowledge) for the first time in the context of automated issue assignment to detect the deteriorations in the assignments made by a data mining model.
    
\subsection{Evaluation}
\label{monitor:eval}

We applied the PELT approach to the daily assignment accuracies collected from the field. PELT detected three change points, each of which was depicted by a vertical dashed line in Figure~\ref{deployment:accuracies}. It turned out that these change points, indeed, coincided with some  important events that affected the assignment accuracies,   validating the results obtained from the proposed approach. The first dashed line represents the date, on which significant changes in the team responsibilities occurred due to migrating certain functionalities from mainframes to state-of-the-art platforms. The time gap between the first and second dashed lines (i.e., about $2.5$ months) represent the amount of time it took for the IT-HD clerks to adapt to these changes. And, the third dashed line represents the date on which IssueTAG was deployed. Further discussion on these change points can be found in Section~\ref{deployment:analysis}.


We observed that PELT did not detect any other change point after IssueTAG was deployed. We believe that this was because the underlying classification model had been regularly retrained at every month as a part of Softtech's policy by using the issue reports submitted in the last 12 months before the calibration (Section~\ref{timeLocality}).

To further evaluate the proposed approach, we, therefore, carried out additional experiments where we systematically varied the nature of the deteriorations and evaluated whether the proposed approach detected them or not. Note that controlling the nature of the deteriorations in this study allows us to reliably evaluate the results, because when the true nature of a deterioration, such as the exact point in time at which the deterioration occurred, is not known, which is typically the case with the data collected from the field, the analysis may suffer from the lack of ground truth. Note further that even if the underlying classification model is regularly trained, monitoring for deteriorations is still relevant as the assignment accuracies can still deteriorate in between the calibrations.

\subsubsection{Experimental Setup}
\label{monitor:setup}

In each experimental setup, we used an ordered sequence of $200$ daily assignment accuracies. The first $100$ of these accuracies came from a normal distribution representing the accuracies expected from IssueTAG, whereas the remaining $100$ accuracies came from a distribution (or a number of distributions) representing a deterioration. That is, the change point in each experiment was the $100$th time point as the deterioration was introduced after this point in time. 

For each experimental setup, we then mimicked the real-life operations of IssueTAG. More specifically, given a sequence of $200$ daily assignment accuracies, we fed them to the proposed approach one daily accuracy after another in the order they appeared in the sequence. After every daily accuracy, a decision was made whether a deterioration had occurred, and if so, when. We finally determined how long it took for the proposed approach to detect the deterioration. For each experimental setup, we repeated the experiments $1000$ times.

As an implementation of the PELT approach, we used the {\tt ruptures} Python library \citep{truong2018ruptures}. As the {\em penalty level}, i.e., the only parameter to calibrate in PELT, we used the empirically determined value of $0.05$. The penalty level is a mechanism used for guarding against overfitting, determining to which extent a shift in the accuracies should be considered as a change point. The larger the penalty level, the fewer (and more significant) change points are detected.



To model the daily accuracies expected from the system,  we used a normal distribution with mean of $0.85$ and standard deviation of $0.025$ (i.e., ${\mu = 0.85}$ and ${\sigma = 0.025}$), mimicking the daily accuracies of the deployed system observed in the field (Section~\ref{deployment:analysis}). To model the deteriorations, we experimented with two types of changes: {\em sudden deteriorations} and {\em gradual deteriorations}. In either case, we used $5$-, $10$-, $15$-, and $20$-point drops in  daily accuracies, such that the mean accuracy (i.e., the mean of the distribution, from which the accuracies were drawn) eventually became $0.80$, $0.75$, $0.70$, and $0.65$, respectively. 

\begin{figure*}
  \includegraphics[width=1\textwidth]{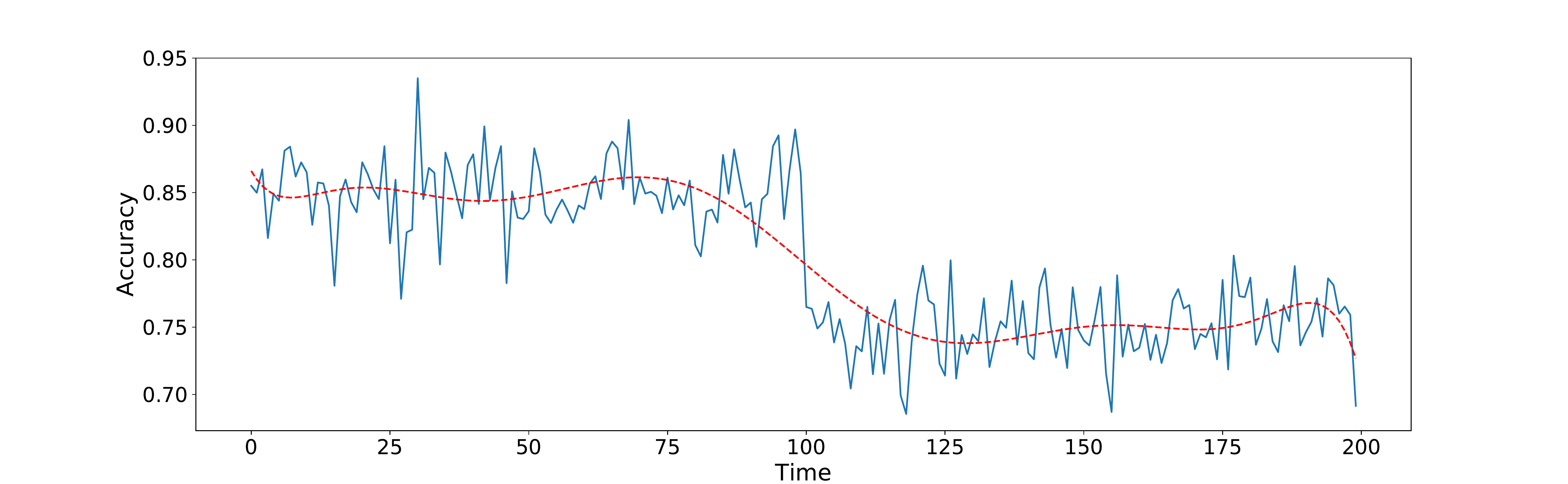}
\caption{An example sequence of daily assignment accuracies showing a sudden $10$-point deterioration at the $100$th time point.}
\label{fig:suddenChange}     
\end{figure*}

For the sudden deteriorations, we abruptly dropped the mean accuracy from $0.85$ to the requested level (i.e., $0.80$, $0.75$, $0.70$, or $0.65$, depending on the choice) right after the change point at the $100$th time point and kept it intact until and including the $200$th time point (i.e., until the end of the experiment). Figure~\ref{fig:suddenChange} presents an example sequence of daily assignment accuracies showing a sudden $10$-point deterioration.

\begin{figure*}
  \includegraphics[width=1\textwidth]{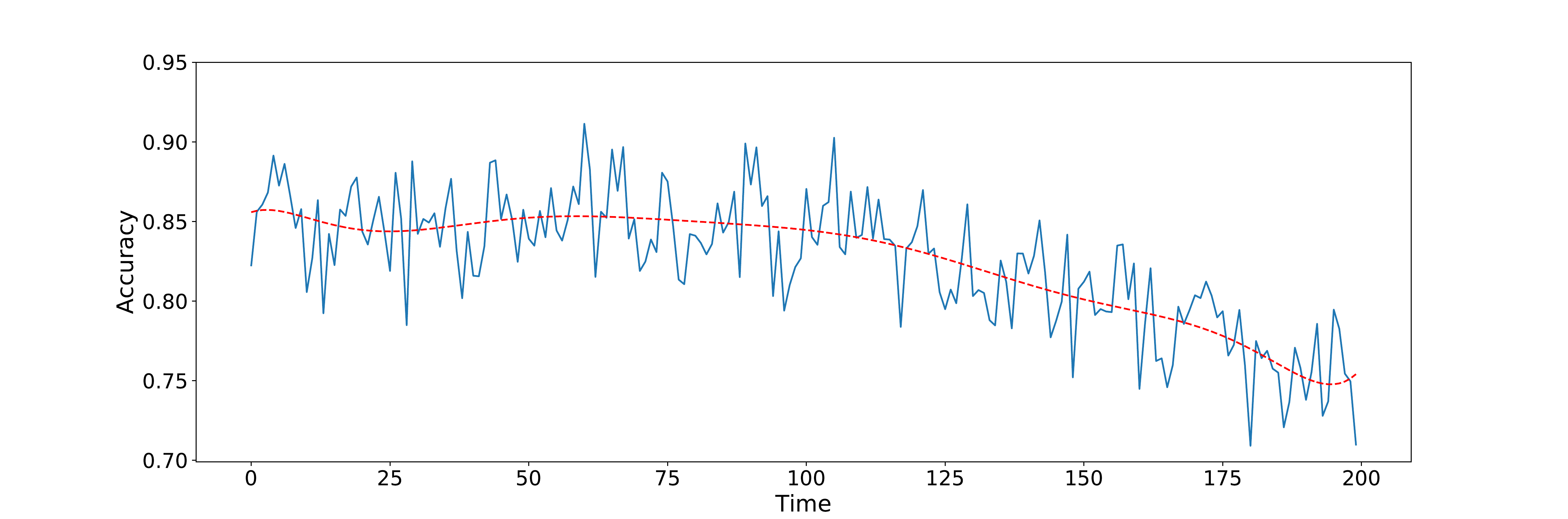}
\caption{An example sequence of daily assignment accuracies showing a gradual $10$-point deterioration starting from the $100$th time point.}
\label{fig:gradualChange}
\end{figure*}

For the gradual deteriorations, on the other hand, the changes were obtained by linearly dropping the mean accuracy starting from right after the change point at the $100$th time point until and including the $200$th time point, such that the mean accuracy at end of the experiment became $0.80$,  $0.75$, $0.70$, or $0.65$, depending on the choice. For example, if the requested level of accuracy was $0.80$, then starting from the mean accuracy of $0.85$, the mean accuracy would be dropped by $0.05$-point each day ($5$-point drop/$100$ days) until it would become $0.80$ at the $200$th time point. Figure~\ref{fig:gradualChange} presents an example sequence of daily assignment accuracies showing a gradual $10$-point deterioration starting from the $100$th time point. 

\subsubsection{Evaluation Framework}
\label{monitor:evalFramework}

To evaluate the proposed approach, we first determine whether the deteriorations are detected or not. If so, we measure {\em detection time} as the number of days past after the change point (i.e., after the $100$th time point) until the deterioration is detected. The smaller the detection time, the better the proposed approach is.

\subsubsection{Data and Analysis}
\label{monitor:analysis}

Table~\ref{tbl:suddendrops} presents the data we obtained on the sudden deteriorations used in the study. We first observed that the proposed approach detected all the  deteriorations. We then observed that as the deterioration amount increased, the detection time tended to decrease, i.e., the proposed approach tended to detect the deteriorations faster. On average, the deteriorations were detected in $1.33$, $1.60$, $1.84$, $2.67$ days after there was a $20$-, $15$-, $10$-, and $5$-point sudden drop in the mean assignment accuracies, respectively.

\begin{table}[b]
\centering
\caption{Results obtained on sudden deteriorations. The experiments were repeated $1000$ times.}
\label{tbl:suddendrops}
\begin{tabular}{rllll}
\hline\noalign{\smallskip}
               & \multicolumn{4}{c}{Detection Time} \\
Deterioration  & min & avg & max & stddev \\
\noalign{\smallskip}\hline\noalign{\smallskip}
5-point  & 1 & 2.67 & 6 & 1.13 \\
10-point & 1 & 1.84 & 3 & 0.38 \\
15-point & 1 & 1.60 & 2 & 0.49 \\
20-point & 1 & 1.33 & 2 & 0.47 \\
\noalign{\smallskip}\hline
\end{tabular}
\end{table}
	
Table~\ref{tbl:gradualdrops} presents the data we obtained on the gradual deteriorations. As was the case with the sudden deteriorations, the proposed approach detected all the deteriorations and as the deterioration amount (thus, the deterioration rate) increased, the detection time tended to decrease. Compared to the sudden deteriorations, however, the detection times for gradual deteriorations increased, which is to be expected. To better evaluate the quality of the detections, we, therefore, analyzed the mean accuracies that were present when the deteriorations were detected. We observed that throughout all the experiments, the proposed approach detected the deteriorations before the mean accuracy dropped more than $5$-points (the last column in Table~\ref{tbl:gradualdrops}).

\begin{table}
\centering
\caption{Results obtained on gradual deteriorations. The experiments were repeated $1000$ times.}
\label{tbl:gradualdrops}    
\begin{tabular}{rllllc}
\hline\noalign{\smallskip}
               & \multicolumn{4}{c}{Detection Time} & Minimum Mean Accuracy  \\
Deterioration  & min & avg & max & stddev & at the Point of Detection\\
\noalign{\smallskip}\hline\noalign{\smallskip}
5-point  & 1 & 31.03 & 55 & 13.64 & 0.8125 \\
10-point & 1 & 20.14 & 35 & 8.41 & 0.8070 \\
15-point & 1 & 15.70 & 26 & 6.49 & 0.8035  \\
20-point & 1 & 13.36 & 21 & 4.95 & 0.8000 \\
\noalign{\smallskip}\hline
\end{tabular}
\end{table}

\section{Lessons Learnt}
\label{lessons}

{\bf Stakeholders do not necessarily resist change.} To deploy IssueTAG, we carried out a number of meetings with the IT-HD, AST, and software development teams. One thing we repeatedly observed in these meetings was that all the stakeholders, although they had some rightful concerns, such as what if the proposed approach adversely affects the issue-resolution process -- a major concern for a company developing business-critical software systems -- were actually willing to automate the process of issue assignments as much as possible. 

We, indeed, had no objection at all. The AST members and the development teams were looking forward to reducing the turnaround time for issue resolutions; the incorrectly assigned issue reports were bouncing back and forth between the IT-HD clerks and the AST members, causing a great deal of wasted time. The IT-HD clerks were looking forward to 1) avoiding the costly and cumbersome process of maintaining a knowledge base about the development teams and their responsibilities and 2) deferring the responsibility of making assignments as much as possible since incorrect assignments were often causing friction with the AST members and the development teams. 

Another reason behind the absence of any resistance was that none of the stakeholders felt threatened by the new system. The IT-HD clerks were still needed as they were the ones communicating with both the bank costumers and employees to collect the issues, resolving the ones that they could, and creating issue reports for the remaining ones. The AST members were still needed as they were the ones helping the development teams manage the issue reports. The development teams were still needed as they were the ones developing the software products.

{\bf Gradual transition helps stakeholders build confidence, facilitating the acceptance of the system.} To address the rightful concerns of the stakeholders regarding the accuracy of the proposed system, we followed a gradual transition strategy. First, we simply added a single button to the screen, which the IT-HD clerks used to create the issue reports. We initially did not modify the assignment process at all in the sense that the use of this button was optional. If the IT-HD clerk chose to arm the button after creating an issue report, it would simply display the assignment made by IssueTAG. The clerk could then accept the assignment as it was or modify it. We observed that $3$ months after the deployment of this button, enough confidence was built among the stakeholders to fully deploy the system.

{\bf It is not just about automating the issue  assignments, but also about changing the process around it.} One observation we made numerous times during the meetings with the stakeholders was that automating the issue assignments also requires to modify the other parts of the assignment process to improve the efficiency and effectiveness of the entire process to the extent possible. This was because most of the steps in the assignment process were dependent on the fact that issue assignments were made by the IT-HD clerks. Changing this, therefore, necessitated other changes. In particular, we prevented the IT-HD clerks from modifying the issue assignments made by IssueTAG and the incorrectly assigned issue reports from being returned back to the IT-HD clerks for a reassignment. All of these changes were based on the discussions we had with the stakeholders as well as the analysis of the results we obtained from a number of feasibility studies (Section~\ref{inaction}).

{\bf The accuracy of the deployed system does not have to be higher than that of manual assignments in order for the system to be useful.} Although the assignment accuracy of IssueTAG was slightly lower than that of manual assignments, it reduced the manual effort required for the assignments and improved the turnaround time for closing the issue reports. All of these helped improve the usability of IssueTAG, which was also evident from the survey we conducted on the stakeholders in the field (Section~\ref{userEvals}).


{\bf Deploying a data mining-based automated issue assignment system requires the development of additional functionalities.} When the issue assignments are automatically made by using a data mining model, the accuracy of the assignments needs to be monitored and deteriorations need to be detected in an online manner, so that corrective actions, such as recalibrating the underlying model, can be taken in time. To this end, we have developed a change point detection-based approach using PELT \citep{killick2012optimal} (Section~\ref{monitor}). Furthermore, stakeholders may demand some explanations as to why certain issue reports (especially the incorrectly assigned ones) have been assigned to their teams. Note that since the data mining models used for predicting the assignments are not necessarily readable and interpretable by human beings (as was the case in this work), generating such explanations can be a non-trivial task. To this end, we have developed a LIME-based \citep{ribeiro2016should} approach for automatically generating explanations that can easily be interpreted even by non-technical stakeholders.

\section{Threats to Validity}
\label{threats}

In this section, we discuss threats to validity.

\subsection{Construct Validity}
\label{constructValidity}

To circumvent the construct threats, we used the well-known {\em accuracy} metric \citep{manning2010introduction} throughout the paper to evaluate the quality of the issue assignments. We have also complemented the accuracy results with other well-known metrics, namely precision, recall, and F-measure, as we see fit (Section~\ref{existing}). We mainly focused on the accuracies because 1) it was the choice of a recent related work in the literature \citep{jonsson2016automated} and 2) the assignment accuracies and F-measures (computed by giving equal importance to both precision and recall) we obtained in the experiments were comparable (Table~\ref{tbl:existing}). 

To measure the amount of effort saved by automating the issue assignments (Section~\ref{inaction}), we used the person-month metric, which is also a well-known metric to quantify effort in software engineering projects \citep{pressman2005software}.

To measure the effect of the proposed approach on the issue-resolution process, we compared the average times required to close the issue reports before and after the deployment of IssueTAG (Section~\ref{inaction}).  To this end, we used the dates and times recorded by the issue report management tool (namely, Jira). Furthermore, since the characteristics of the reported issues, thus the times it takes to resolve them, can change over time, we used the issue reports submitted within two months before and after the deployment of the system for this purpose.
 
To further evaluate the usefulness of the deployed system, we carried out a survey on the actual users of the system (Sections~\ref{userEvals}-\ref{explain}). The survey had both Likert scale and open-ended questions and about half of the actual users of the deployed system voluntarily participated in the survey.

Throughout the paper, we used the actual database of issue reports maintained by Softtech. Furthermore, all the survey results were obtained from the actual users in the field. We followed the same approach to evaluate our PELT-based technique to detect deteriorations in assignment accuracies, which, indeed, successfully detected three deteriorations each with a different cause (Section~\ref{monitor}). To further evaluate the proposed approach, we also carried out controlled experiments, each of which was repeated $1000$ times. We did this because in the data collected from the field, it was not always possible to determine whether there really were some deteriorations or not, and if so, what the nature of these deteriorations were. Therefore, the controlled experiments helped us further evaluate the proposed approach, as in these experiments, we knew both the nature of the deteriorations (e.g., sudden or gradual) and the exact point in time where they occurred.

\subsection{Internal Validity}
\label{internalValidity}


To circumvent the internal threats that may be caused by implementation errors, we used well-known and frequently used libraries. In particular, we used the Python scikit-learn \citep{pedregosa2011scikit} library for preprocessing the issue reports and extracting the features; the scikit-learn \citep{pedregosa2011scikit} and mlxtend \citep{raschka2018mlxtend} libraries for training the classification models; the lime \citep{ribeiro2016should} library for creating the LIME-based explanations for the assignments; and the ruptures \citep{truong2018ruptures} library for PELT-based change point detection.

In Section~\ref{existing}, we performed the same preprocessing steps and extracted the same set of features for all the classification algorithms used in the study. However, the performances of these classifiers might have been dependent on the preprocessing steps used and the features extracted. On the other hand, we used well-known preprocessing steps, such as tokenization and removal of non-letter characters as well as stop words and extracted frequently used features, such as the bag-of-words model.

A related concern is that we used the default configurations of the aforementioned classifiers, except for the $k$-nearest neighbor and the stacked generalization classifiers where we used cosine similarity and empirically tuned $k$ for the former; and logistic regression as the level-1 algorithm together with the probabilities emitted by the level-0 classifiers for the latter. On the other hand, the performance of these classifiers might have been dependent on the underlying configurations. Note, however, that optimizing the configurations for these classifiers could have only generated better accuracies. 

In the evaluations, as the correct team for a given issue report (i.e., as the ground truth), we used the team who actually closed the report. Some reports, however, might have needed to be processed by multiple teams before the reported issues could be fixed. Since in these situations, typically the last team in the chain closed the report, even if the initial assignment of the report was considered to be correct, it was counted as incorrect when computing the assignment accuracies. Note, however, that counting such assignments as correct could have only increased the accuracies.

When computing the amount of manual effort required for issue assignments, we did not take the amount of effort required for maintaining the knowledge base used by the IT-HD clerks into account. Therefore, the actual savings in person-months can be larger than the ones reported.

\subsection{External Validity}
\label{externalValidity}


One external threat is that IssueTAG was deployed at Softtech/IsBank only. Softtech, however, being a subsidiary of IsBank -- the largest private bank in Turkey -- is the largest software company of Turkey owned by domestic capital, maintaining around $100$ millions of lines of code with $1.200$ employees. Consequently, it shares many characteristics of large software development houses, especially the ones producing custom, business-critical software systems, such as having a large, continuously evolving code base maintained by dozens of development teams involving hundreds of software developers with hundreds of issue reports filed daily, each of which needs to be  addressed with utmost importance and urgency.

Another possible threat is that issue reports at IsBank (thus, the ones used in this work) are created by the IT-HD clerks  (Section~\ref{caseDescription}). Although, this team is a non-technical team, they are specialized in creating issue reports by listening to the bank customers and employees. Therefore, the quality of the issue reports used in this study may differ from the ones directly created by, for example, the end-users of a system. However, many companies, especially the ones that produce business-critical software systems and that need to deal with a large number of issue reports, employ similar call centers. Furthermore, all the issue reports used in this work were written in Turkish. However, we used simple text processing steps, such as tokenization and removal of non-letter characters and stop words. Therefore, the proposed approach can also be used with issue reports written in other languages.

\subsection{Conclusion Validity}
\label{conclusionValidity}

All the issue reports we used in the experiments were the real issue reports collected from the field. After the deployment of IssueTAG, once an issue report was created by an IT-HD clerk, the assignment was automatically made by the system. There was no means that the deployed system could be bypassed or that the assignments made by the system could be changed by an IT-HD clerk. Note that the AST members could then reassign the issue reports if needed, in which case the initial assignments made by the system were considered as incorrect. The number of issue reports closed was an important performance metric for the AST members as well as for the development teams at Softtech. Consequently, as a part of the company's policy, the issue reports were required to be closed by the development teams, who actually resolved the reported issues. The stakeholders payed utmost attention to this matter.  Therefore, the assignment accuracies reported in this work, reflect the actual accuracies obtained by IssueTAG in the field.

To further evaluate the deployed system, we carried out two surveys (Sections~\ref{userEvals}-\ref{explain}). Although $14$ participants were involved in these surveys, these participants constituted about half ($14$ out of $30$) of the AST members, who are the direct end-users of IssueTAG, dealing with the issue reports on a daily basis.

\section{Related Work}
\label{related}

Several works in the literature studied the issue assignment problem. These works use a variety of approaches to make the assignments, including Naive Bayes classifiers \citep{murphy2004automatic, anvik2006should}, Bayesian Networks \citep{jeong2009improving}, Support Vector Machines \citep{anvik2006should, jonsson2016automated}, and information retrieval-based approaches \citep{chen2011approach, kagdi2012assigning, nagwani2012predicting, shokripour2012automatic, canfora2006supporting, linares2012triaging, xie2012dretom, xia2013accurate}, Expectation Maximization \citep{anvik2007assisting}, Nearest Neighbor classifiers \citep{anvik2011reducing}, Decision Trees \citep{ahsan2009automatic}, Random Forests \citep{ahsan2009automatic}, REPTrees \citep{ahsan2009automatic}, Radial Basis Function Networks \citep{ahsan2009automatic}, Neural Networks \citep{helming2010automatic} and Ensemble-based classification \citep{jonsson2016automated}.

These works, except for \citep{lin2009empirical, helming2010automatic, jonsson2016automated, dedik2016automated}, evaluated the proposed approaches by using the issue databases of open source projects. We, on the other hand, used the issue reports filed for commercial, closed-source projects. Although the remaining works \citep{lin2009empirical, helming2010automatic, jonsson2016automated, dedik2016automated}, report on the results obtained on closed-source, commercial software projects, they do so by carrying out a retrospective analysis in an offline manner. We, on the other hand, deployed the proposed approach and shared both the results we obtained and the lessons we learnt regarding the practical effects of automated issue assignment in the field. Furthermore, to the best of our knowledge, our work is the first work carrying out user studies in this context.

Some of the aforementioned works use natural language explanations present in issue reports for assignments, such as one-line summary and description \citep{murphy2004automatic, anvik2006should, canfora2006supporting, ahsan2009automatic, baysal2009bug, jeong2009improving, lin2009empirical, matter2009assigning, helming2010automatic, anvik2011reducing, chen2011approach, park2011costriage, bhattacharya2012automated, linares2012triaging, nagwani2012predicting, alenezi2013efficient, jonsson2016automated, bettenburg2008makes}. Others also leverage categorical information, such as product, component, and  version \citep{ahsan2009automatic, lin2009empirical, park2011costriage, jonsson2016automated}. 

In this work, we used natural language descriptions present in the issue reports, more specifically the one-line summaries and descriptions. We did not use any categorical information, e.g., product, component, and version information, because such information was not included in the issue reports; there were no fields in the issue reporting tool, requesting these types of categorical information.  The reason was that with the collection of software products maintained by Softtech, which heavily interact with each other in a  business-critical environment, sharing many resources, such as databases, file systems, and GUI screens, the boundaries of the products from the perspective of issue reporting were not clear at all. Further discussion on this can be found in Section~\ref{caseDescription}.

Different sources of information have been also used for making the assignments. For example, \citep{tamrawi2011fuzzy} model the technical expertise of individual developers and use these models together with the information about the developers who recently made changes in the code base. \citep{wu2011drex} infer a social network model of the developers using the comments they make on historical issue reports as well as the comments automatically generated at the time of the source code commits, to help with the assignments. \citep{baysal2009bug} use developers' preferences as an additional source of information, which are expressed by the ratings the developers gave for the issues they resolved.

We, in this work, deliberately used a single source of information, i.e., the natural language descriptions present in the issue reports, to simplify the design and implementation of the proposed system to the extent possible. This was a design decision we made to increase the reliability of the proposed system as the system needed to be deployed, making hundreds of assignments per day in a business-critical environment. However, we are currently in the process of figuring out what types of additional sources of information could be used in an industrial setup to further improve the assignment accuracies. 



There are also automated approaches for dealing with various other aspects of the issue report management process. One type of approaches aim to identify duplicate issue reports, which can help developers with 1) figuring out the number of actual issues reported; 2) assigning priorities; and 3) debugging \citep{podgurski}. Generally speaking the problem of duplicate identification is casted to a clustering problem where similar reports are grouped together with the assumption that similar descriptions report the same (or similar) issues \citep{podgurski, bettenburg2008duplicate, wang2008approach, jalbert2008automated}.


Other types of approaches mainly focus on better utilizing the available resources for resolving the reported issues. For example, some approaches aim to predict the severities of the issues \citep{lamkanfi2010predicting, menzies2008automated, antoniol2008bug, pandey2017automated}, which, in this context, indicate the levels of impact the issues have on the development and release process. Others aim to predict the effort required to resolve the issues \citep{weiss2007long, giger2010predicting, zhang2013predicting}.



Note that the aforementioned problems, i.e., duplicate detection, severity identification, and effort prediction are different than the issue assignment problem addressed in this work. We, however, plan to conduct industrial-strength studies at IsBank and Softtech to evaluate the efficiency and effectiveness of these approaches.

\section{Conclusion and Future Work}
\label{conclusion}

In this work, we have developed and deployed a system to automate the process of issue assignments at Softtech/IsBank. To this end, we first cast the problem to a classification problem and determined the classifier to be used in the deployed system by empirically evaluating a number of existing classifiers, which are known to perform well for the problem at hand, on the actual database of issues maintained by the company. We then carried out further studies to determine both the amount and time locality of the historical data required for training the underlying classification models. We finally deployed the proposed system by configuring it based on the results we obtained from these studies.

We observed that 1) it is not just about deploying a data mining-based system for automated issue assignment, but also about designing/changing the assignment process around the system to get the most out of it; 2) the accuracy of the system does not have to be higher than that of manual assignments in order for the system to be useful, which was further validated by the user studies we carried out on actual stakeholders in the field; 3) deploying such a system  also requires the development of additional functionalities, such as detecting deteriorations in assignment accuracies in an online manner and creating human-readable, non-technical explanations for the assignments made, for both of which we developed and empirically evaluated different approaches; 4) stakeholders do not necessarily resist change; and 5) gradual transitions can help stakeholders build confidence, which, in turn, facilitates the acceptance of the system.
 
One avenue for future research is to use additional sources of information to further improve the assignment accuracy; we are, in particular, interested in developing ``human-in-the-loop'' type of approaches. Another avenue is to carry out industrial-strength studies using the deployed system to evaluate the efficiency and effectiveness of the other related approaches in the field, including duplicate detection, severity identification, and effort prediction.




\begin{thebibliography}{53}
\providecommand{\natexlab}[1]{#1}
\providecommand{\url}[1]{{#1}}
\providecommand{\urlprefix}{URL }
\expandafter\ifx\csname urlstyle\endcsname\relax
  \providecommand{\doi}[1]{DOI~\discretionary{}{}{}#1}\else
  \providecommand{\doi}{DOI~\discretionary{}{}{}\begingroup
  \urlstyle{rm}\Url}\fi
\providecommand{\eprint}[2][]{\url{#2}}

\bibitem[{Ahsan et~al.(2009)Ahsan, Ferzund, and Wotawa}]{ahsan2009automatic}
Ahsan SN, Ferzund J, Wotawa F (2009) Automatic software bug triage system (bts)
  based on latent semantic indexing and support vector machine. In: 2009 Fourth
  International Conference on Software Engineering Advances, IEEE, pp 216--221

\bibitem[{Alenezi et~al.(2013)Alenezi, Magel, and
  Banitaan}]{alenezi2013efficient}
Alenezi M, Magel K, Banitaan S (2013) Efficient bug triaging using text mining.
  JSW 8(9):2185--2190

\bibitem[{Antoniol et~al.(2008)Antoniol, Ayari, Di~Penta, Khomh, and
  Gu{\'e}h{\'e}neuc}]{antoniol2008bug}
Antoniol G, Ayari K, Di~Penta M, Khomh F, Gu{\'e}h{\'e}neuc YG (2008) Is it a
  bug or an enhancement?: a text-based approach to classify change requests.
  In: CASCON, vol~8, pp 304--318

\bibitem[{Anvik(2007)}]{anvik2007assisting}
Anvik J (2007) Assisting bug report triage through recommendation. PhD thesis,
  University of British Columbia

\bibitem[{Anvik and Murphy(2011)}]{anvik2011reducing}
Anvik J, Murphy GC (2011) Reducing the effort of bug report triage:
  Recommenders for development-oriented decisions. ACM Transactions on Software
  Engineering and Methodology (TOSEM) 20(3):10

\bibitem[{Anvik et~al.(2006)Anvik, Hiew, and Murphy}]{anvik2006should}
Anvik J, Hiew L, Murphy GC (2006) Who should fix this bug? In: Proceedings of
  the 28th international conference on Software engineering, ACM, pp 361--370

\bibitem[{Baysal et~al.(2009)Baysal, Godfrey, and Cohen}]{baysal2009bug}
Baysal O, Godfrey MW, Cohen R (2009) A bug you like: A framework for automated
  assignment of bugs. In: 2009 IEEE 17th International Conference on Program
  Comprehension, IEEE, pp 297--298

\bibitem[{Bettenburg et~al.(2008{\natexlab{a}})Bettenburg, Just, Schr{\"o}ter,
  Weiss, Premraj, and Zimmermann}]{bettenburg2008makes}
Bettenburg N, Just S, Schr{\"o}ter A, Weiss C, Premraj R, Zimmermann T
  (2008{\natexlab{a}}) What makes a good bug report? In: Proceedings of the
  16th ACM SIGSOFT International Symposium on Foundations of software
  engineering, ACM, pp 308--318

\bibitem[{Bettenburg et~al.(2008{\natexlab{b}})Bettenburg, Premraj, Zimmermann,
  and Kim}]{bettenburg2008duplicate}
Bettenburg N, Premraj R, Zimmermann T, Kim S (2008{\natexlab{b}}) Duplicate bug
  reports considered harmful… really? In: 2008 IEEE International Conference
  on Software Maintenance, IEEE, pp 337--345

\bibitem[{Bhattacharya et~al.(2012)Bhattacharya, Neamtiu, and
  Shelton}]{bhattacharya2012automated}
Bhattacharya P, Neamtiu I, Shelton CR (2012) Automated, highly-accurate, bug
  assignment using machine learning and tossing graphs. Journal of Systems and
  Software 85(10):2275--2292

\bibitem[{Bishop(2006)}]{bishop2006pattern}
Bishop CM (2006) Pattern recognition and machine learning. springer

\bibitem[{Breiman(2001)}]{breiman2001random}
Breiman L (2001) Random forests. Machine Learning 45(1):5--32

\bibitem[{Breiman(2017)}]{breiman2017classification}
Breiman L (2017) Classification and regression trees. Routledge

\bibitem[{Canfora and Cerulo(2006)}]{canfora2006supporting}
Canfora G, Cerulo L (2006) Supporting change request assignment in open source
  development. In: Proceedings of the 2006 ACM symposium on Applied computing,
  ACM, pp 1767--1772

\bibitem[{Chen et~al.(2011)Chen, Wang, and Liu}]{chen2011approach}
Chen L, Wang X, Liu C (2011) An approach to improving bug assignment with bug
  tossing graphs and bug similarities. JSW 6(3):421--427

\bibitem[{Ded{\'\i}k and Rossi(2016)}]{dedik2016automated}
Ded{\'\i}k V, Rossi B (2016) Automated bug triaging in an industrial context.
  In: 2016 42th Euromicro Conference on Software Engineering and Advanced
  Applications (SEAA), IEEE, pp 363--367

\bibitem[{Giger et~al.(2010)Giger, Pinzger, and Gall}]{giger2010predicting}
Giger E, Pinzger M, Gall H (2010) Predicting the fix time of bugs. In:
  Proceedings of the 2nd International Workshop on Recommendation Systems for
  Software Engineering, ACM, pp 52--56

\bibitem[{Helming et~al.(2010)Helming, Arndt, Hodaie, Koegel, and
  Narayan}]{helming2010automatic}
Helming J, Arndt H, Hodaie Z, Koegel M, Narayan N (2010) Automatic assignment
  of work items. In: International Conference on Evaluation of Novel Approaches
  to Software Engineering, Springer, pp 236--250

\bibitem[{Hocking et~al.(2013)Hocking, Schleiermacher, Janoueix-Lerosey,
  Delattre, Bach, and Vert}]{hocking2013learning}
Hocking TD, Schleiermacher G, Janoueix-Lerosey I, Delattre O, Bach F, Vert JP
  (2013) Learning smoothing models of copy number profiles using breakpoint
  annotations. BMC bioinformatics 14:164

\bibitem[{Jalbert and Weimer(2008)}]{jalbert2008automated}
Jalbert N, Weimer W (2008) Automated duplicate detection for bug tracking
  systems. In: 2008 IEEE International Conference on Dependable Systems and
  Networks With FTCS and DCC (DSN), IEEE, pp 52--61

\bibitem[{Jeong et~al.(2009)Jeong, Kim, and Zimmermann}]{jeong2009improving}
Jeong G, Kim S, Zimmermann T (2009) Improving bug triage with bug tossing
  graphs. In: Proceedings of the the 7th joint meeting of the European software
  engineering conference and the ACM SIGSOFT symposium on The foundations of
  software engineering, ACM, pp 111--120

\bibitem[{Joachims(1998)}]{joachims1998text}
Joachims T (1998) Text categorization with support vector machines: Learning
  with many relevant features. In: Proceedings of the 10th European Conference
  on Machine Learning, Springer-Verlag, ECML'98, pp 137--142

\bibitem[{Jonsson et~al.(2016)Jonsson, Borg, Broman, Sandahl, Eldh, and
  Runeson}]{jonsson2016automated}
Jonsson L, Borg M, Broman D, Sandahl K, Eldh S, Runeson P (2016) Automated bug
  assignment: Ensemble-based machine learning in large scale industrial
  contexts. Empirical Software Engineering 21(4):1533--1578

\bibitem[{Kagdi et~al.(2012)Kagdi, Gethers, Poshyvanyk, and
  Hammad}]{kagdi2012assigning}
Kagdi H, Gethers M, Poshyvanyk D, Hammad M (2012) Assigning change requests to
  software developers. Journal of Software: Evolution and Process 24(1):3--33

\bibitem[{Killick et~al.(2012)Killick, Fearnhead, and
  Eckley}]{killick2012optimal}
Killick R, Fearnhead P, Eckley IA (2012) Optimal detection of changepoints with
  a linear computational cost. Journal of the American Statistical Association
  107(500):1590--1598

\bibitem[{Lamkanfi et~al.(2010)Lamkanfi, Demeyer, Giger, and
  Goethals}]{lamkanfi2010predicting}
Lamkanfi A, Demeyer S, Giger E, Goethals B (2010) Predicting the severity of a
  reported bug. In: 2010 7th IEEE Working Conference on Mining Software
  Repositories (MSR 2010), IEEE, pp 1--10

\bibitem[{Lavielle and Ere(2007)}]{lavielle2007adaptive}
Lavielle M, Ere G (2007) Adaptive detection of multiple change-points in asset
  price volatility. Long Memory in Economics

\bibitem[{Lin et~al.(2009)Lin, Shu, Yang, Hu, and Wang}]{lin2009empirical}
Lin Z, Shu F, Yang Y, Hu C, Wang Q (2009) An empirical study on bug assignment
  automation using chinese bug data. In: 2009 3rd International Symposium on
  Empirical Software Engineering and Measurement, IEEE, pp 451--455

\bibitem[{Linares-V{\'a}squez et~al.(2012)Linares-V{\'a}squez, Hossen, Dang,
  Kagdi, Gethers, and Poshyvanyk}]{linares2012triaging}
Linares-V{\'a}squez M, Hossen K, Dang H, Kagdi H, Gethers M, Poshyvanyk D
  (2012) Triaging incoming change requests: Bug or commit history, or code
  authorship? In: 2012 28th IEEE International Conference on Software
  Maintenance (ICSM), IEEE, pp 451--460

\bibitem[{Manning et~al.(2010)Manning, Raghavan, and
  Sch{\"u}tze}]{manning2010introduction}
Manning C, Raghavan P, Sch{\"u}tze H (2010) Introduction to information
  retrieval. Natural Language Engineering 16(1):100--103

\bibitem[{Matter et~al.(2009)Matter, Kuhn, and
  Nierstrasz}]{matter2009assigning}
Matter D, Kuhn A, Nierstrasz O (2009) Assigning bug reports using a
  vocabulary-based expertise model of developers. In: 2009 6th IEEE
  international working conference on mining software repositories, IEEE, pp
  131--140

\bibitem[{Menzies and Marcus(2008)}]{menzies2008automated}
Menzies T, Marcus A (2008) Automated severity assessment of software defect
  reports. In: 2008 IEEE International Conference on Software Maintenance,
  IEEE, pp 346--355

\bibitem[{Murphy and Cubranic(2004)}]{murphy2004automatic}
Murphy G, Cubranic D (2004) Automatic bug triage using text categorization. In:
  Proceedings of the Sixteenth International Conference on Software Engineering
  \& Knowledge Engineering, Citeseer

\bibitem[{Nagwani and Verma(2012)}]{nagwani2012predicting}
Nagwani NK, Verma S (2012) Predicting expert developers for newly reported bugs
  using frequent terms similarities of bug attributes. In: 2011 Ninth
  International Conference on ICT and Knowledge Engineering, IEEE, pp 113--117

\bibitem[{Pandey et~al.(2017)Pandey, Sanyal, Hudait, and
  Sen}]{pandey2017automated}
Pandey N, Sanyal DK, Hudait A, Sen A (2017) Automated classification of
  software issue reports using machine learning techniques: an empirical study.
  Innovations in Systems and Software Engineering 13(4):279--297

\bibitem[{Park et~al.(2011)Park, Lee, Kim, Hwang, and Kim}]{park2011costriage}
Park Jw, Lee MW, Kim J, Hwang Sw, Kim S (2011) Costriage: A cost-aware triage
  algorithm for bug reporting systems. In: Twenty-Fifth AAAI Conference on
  Artificial Intelligence

\bibitem[{Pedregosa et~al.(2011)Pedregosa, Varoquaux, Gramfort, Michel,
  Thirion, Grisel, Blondel, Prettenhofer, Weiss, Dubourg
  et~al.}]{pedregosa2011scikit}
Pedregosa F, Varoquaux G, Gramfort A, Michel V, Thirion B, Grisel O, Blondel M,
  Prettenhofer P, Weiss R, Dubourg V, et~al. (2011) Scikit-learn: Machine
  learning in python. Journal of machine learning research 12(Oct):2825--2830

\bibitem[{{Podgurski} et~al.(2003){Podgurski}, {Leon}, {Francis}, {Masri},
  {Minch}, {Jiayang Sun}, and {Bin Wang}}]{podgurski}
{Podgurski} A, {Leon} D, {Francis} P, {Masri} W, {Minch} M, {Jiayang Sun}, {Bin
  Wang} (2003) Automated support for classifying software failure reports. In:
  25th International Conference on Software Engineering, 2003. Proceedings., pp
  465--475

\bibitem[{Pressman(2005)}]{pressman2005software}
Pressman RS (2005) Software engineering: a practitioner's approach. Palgrave
  Macmillan

\bibitem[{Raschka(2018)}]{raschka2018mlxtend}
Raschka S (2018) Mlxtend: Providing machine learning and data science utilities
  and extensions to python's scientific computing stack. J Open Source Software
  3(24):638

\bibitem[{Ribeiro et~al.(2016)Ribeiro, Singh, and Guestrin}]{ribeiro2016should}
Ribeiro MT, Singh S, Guestrin C (2016) Why should i trust you?: Explaining the
  predictions of any classifier. In: Proceedings of the 22nd ACM SIGKDD
  international conference on knowledge discovery and data mining, ACM, pp
  1135--1144

\bibitem[{Shokripour et~al.(2012)Shokripour, Kasirun, Zamani, and
  Anvik}]{shokripour2012automatic}
Shokripour R, Kasirun ZM, Zamani S, Anvik J (2012) Automatic bug assignment
  using information extraction methods. In: 2012 International Conference on
  Advanced Computer Science Applications and Technologies (ACSAT), IEEE, pp
  144--149

\bibitem[{Tamrawi et~al.(2011)Tamrawi, Nguyen, Al-Kofahi, and
  Nguyen}]{tamrawi2011fuzzy}
Tamrawi A, Nguyen TT, Al-Kofahi JM, Nguyen TN (2011) Fuzzy set and cache-based
  approach for bug triaging. In: Proceedings of the 19th ACM SIGSOFT symposium
  and the 13th European conference on Foundations of software engineering, ACM,
  pp 365--375

\bibitem[{Ting and Witten(1999)}]{ting1999issues}
Ting KM, Witten IH (1999) Issues in stacked generalization. Journal of
  artificial intelligence research 10:271--289

\bibitem[{Truong et~al.(2018{\natexlab{a}})Truong, Oudre, and
  Vayatis}]{truong2018ruptures}
Truong C, Oudre L, Vayatis N (2018{\natexlab{a}}) ruptures: change point
  detection in python. arXiv preprint arXiv:180100826

\bibitem[{Truong et~al.(2018{\natexlab{b}})Truong, Oudre, and
  Vayatis}]{truong2018selective}
Truong C, Oudre L, Vayatis N (2018{\natexlab{b}}) Selective review of offline
  change point detection methods. arXiv preprint arXiv:180100718

\bibitem[{Wang et~al.(2008)Wang, Zhang, Xie, Anvik, and Sun}]{wang2008approach}
Wang X, Zhang L, Xie T, Anvik J, Sun J (2008) An approach to detecting
  duplicate bug reports using natural language and execution information. In:
  Proceedings of the 30th international conference on Software engineering,
  ACM, pp 461--470

\bibitem[{Weiss et~al.(2007)Weiss, Premraj, Zimmermann, and
  Zeller}]{weiss2007long}
Weiss C, Premraj R, Zimmermann T, Zeller A (2007) How long will it take to fix
  this bug? In: Fourth International Workshop on Mining Software Repositories
  (MSR'07: ICSE Workshops 2007), IEEE, pp 1--1

\bibitem[{Wolpert(1992)}]{wolpert1992stacked}
Wolpert DH (1992) Stacked generalization. Neural networks 5(2):241--259

\bibitem[{Wu et~al.(2011)Wu, Zhang, Yang, and Wang}]{wu2011drex}
Wu W, Zhang W, Yang Y, Wang Q (2011) Drex: Developer recommendation with
  k-nearest-neighbor search and expertise ranking. In: 2011 18th Asia-Pacific
  Software Engineering Conference, IEEE, pp 389--396

\bibitem[{Xia et~al.(2013)Xia, Lo, Wang, and Zhou}]{xia2013accurate}
Xia X, Lo D, Wang X, Zhou B (2013) Accurate developer recommendation for bug
  resolution. In: 2013 20th Working Conference on Reverse Engineering (WCRE),
  IEEE, pp 72--81

\bibitem[{Xie et~al.(2012)Xie, Zhang, Yang, and Wang}]{xie2012dretom}
Xie X, Zhang W, Yang Y, Wang Q (2012) Dretom: Developer recommendation based on
  topic models for bug resolution. In: Proceedings of the 8th international
  conference on predictive models in software engineering, ACM, pp 19--28

\bibitem[{Zhang et~al.(2013)Zhang, Gong, and Versteeg}]{zhang2013predicting}
Zhang H, Gong L, Versteeg S (2013) Predicting bug-fixing time: an empirical
  study of commercial software projects. In: Proceedings of the 2013
  international conference on software engineering, IEEE Press, pp 1042--1051

\end{thebibliography}

%
%

\end{document}